\documentclass[a4paper,11pt]{article}

\usepackage{jheppub}                
\usepackage[all]{hypcap}            
\usepackage{slashed}                
\usepackage[usenames,table]{xcolor} 
\usepackage{feyndiag}               
\usepackage{parskip}                
\usepackage{multirow}               
\usepackage{array}                  
\usepackage[section]{placeins}      
\usepackage[utf8]{inputenc}         

\graphicspath{{./figures/} {./plots/}}
\title{Hunting for Dark Matter Coannihilation by Mixing Dijet Resonances and Missing Transverse Energy}
\author[a]{Malte~Buschmann,}
\author[a]{Sonia~El~Hedri,}
\author[a]{Anna~Kaminska,}
\author[a]{Jia~Liu,}
\author[a]{Maikel~de~Vries,}
\author[a]{Xiao-Ping~Wang,}
\author[a]{Felix~Yu,}
\author[b,c]{Jos\'e~Zurita}
\affiliation[a]{PRISMA Cluster of Excellence \& Mainz Institute for Theoretical Physics, Johannes Gutenberg University, 55099 Mainz, Germany}
\affiliation[b]{Institute for Theoretical Physics (ITP), Karlsruhe Institute of Technology, Engesserstra{\ss}e 7, D-76128 Karlsruhe, Germany}
\affiliation[c]{Institute for Nuclear Physics (IKP), Karlsruhe Institute of Technology, Hermann-von-Helmholtz-Platz 1, D-76344 Eggenstein-Leopoldshafen, Germany}
\emailAdd{buschman@uni-mainz.de}
\emailAdd{elhed001@uni-mainz.de}
\emailAdd{akaminsk@uni-mainz.de}
\emailAdd{liuj@uni-mainz.de}
\emailAdd{mdevrie@uni-mainz.de}
\emailAdd{xiaowang@uni-mainz.de}
\emailAdd{yu001@uni-mainz.de}
\emailAdd{jose.zurita@kit.edu}
\preprint{MITP/16-039 \\ \vspace{-8mm} \begin{flushright} TTP/16-019 \end{flushright}}

\abstract{Simplified models of the dark matter (co)annihilation mechanism predict striking new collider signatures untested by current searches.  These models, which were codified in the coannihilation codex, provide the basis for a dark matter (DM) discovery program at the Large Hadron Collider (LHC) driven by the measured DM relic density.  In this work, we study an exemplary model featuring $s$-channel DM coannihilation through a scalar diquark mediator as a representative case study of scenarios with strongly interacting coannihilation partners.  We discuss the full phenomenology of the model, ranging from low energy flavor constraints, vacuum stability requirements, and precision Higgs effects to direct detection and indirect detection prospects.  Moreover, motivated by the relic density calculation, we find significant portions of parameter space are compatible with current collider constraints and can be probed by future searches, including a proposed analysis for the novel signature of a dijet resonance accompanied by missing transverse energy (MET).  Our results show that the $13$~TeV LHC with $100~\mathrm{fb}^{-1}$ luminosity should be sensitive to mediators as heavy as $1$~TeV and dark matter in the 400--500 GeV range.  The combination of searches for single and paired dijet peaks, non-resonant jets + MET excesses, and our novel resonant dijet + MET signature have strong coverage of the motivated relic density region, reflecting the tight connections between particles determining the dark matter abundance and their experimental signatures at the LHC.
}

\begin{document}

\maketitle
\clearpage

\section{Introduction}
\label{sec:introduction}
With the resumption of the Large Hadron Collider (LHC) operating at $\sqrt{s} = 13$~TeV, the hunt for collider signals of dark matter (DM) production has again begun in earnest.  A positive signal of dark matter at the LHC would spur a revolution in particle physics and astrophysics, although the huge breadth of dark matter models and their concomitant collider signatures makes designing a search strategy a daunting task.  In particular, most of our current knowledge voids possible interactions of the DM, rendering it colorless and electrically neutral, while the only concrete DM measurement is its relic density, $\Omega h^2 = 0.1198 \pm 0.0026$~\cite{Ade:2015xua, Agashe:2014kda}.

Nevertheless, the relic density requirement has not featured prominently in collider searches thus far.  In particular, the dark matter relic density can be driven by all interactions of dark sector particles, while collider searches based on effective operators~\cite{Beltran:2010ww, Goodman:2010yf, Bai:2010hh, Goodman:2010ku, Bartels:2012ui} or dark matter pair annihilation simplified models~\cite{Alves:2011wf, An:2013xka, DiFranzo:2013vra, deSimone:2014pda, Abdallah:2014hon, Buckley:2014fba, Harris:2014hga, Garny:2015wea, Abdallah:2015ter} eschew such complications.  As a result, the collider searches for dark matter are mainly variations on a theme of missing transverse energy~\cite{Khachatryan:2014rra, Khachatryan:2014rwa, Aad:2014tda, Aad:2015zva, CMS:2015jdt, Abercrombie:2015wmb} instead of being driven by the known dark matter relic density.

In reference~\cite{Baker:2015qna}, we established a bottom-up framework for dark matter discovery at the LHC based on a simplified model treatment of the (co)annihilation mechanism of thermal relic dark matter.  This approach uncovered several new signatures ripe for analysis, and more importantly, ensures a direct connection between the relic density calculation and the collider signatures associated with the model.  One novelty with this construction was the inclusion of DM coannihilation~\cite{Griest:1990kh} in addition to pair annihilation, as the mere presence of the coannihilation partner together with the mediator radically adds to the complexity and variety of DM collider signatures.  We stress that, unlike the dark matter field, the coannihilating partner and the mediator of the coannihilation diagram can have color and electromagnetic charges.  In this way, the collider signatures of the dark matter, coannihilation partner, and the mediator are driven by production and decay modes dictated by the coannihilation codex~\cite{Baker:2015qna}.

Because the relic density measurement motivates the simplified model construction, we use the $\Omega h^2$ calculation to inform the most promising parameter space for a dark matter discovery at the LHC.  In this way, naive expectations about DM phenomenology from the weakly-interacting massive particle (WIMP) miracle (see reference~\cite{Bertone:2004pz} for a review) are sharpened into concrete predictions for dark sector particle masses and couplings with real discovery prospects at the LHC.  Our approach also complements the broader community efforts at exploring non-WIMP phenomenology at colliders (see, e.g., references~\cite{Feng:2008ya, Feng:2009mn, Bell:2013wua, Hochberg:2014dra, Izaguirre:2015zva}).  

In this work, we continue our exploration of the models presented in the codex, again focusing on the case when the coannihilation partner and the $s$-channel mediator are colored particles.  In contrast with the $s$-channel leptoquark case study presented in reference~\cite{Baker:2015qna}, though, we study an $s$-channel diquark mediator with Yukawa couplings to first generation quarks.  Since the dark matter pair annihilation rate is suppressed, the DM relic density will mainly be set by processes involving only strong interactions.  As a result, the dark sector mass scale can be readily estimated for a wide range of strongly interacting models which depend only on the $SU(3)_C$ representation of the coannihilation partner and its relative splitting with the DM particle. For splittings of about ${\cal O} (1-10)\%$ of the dark matter mass, this scale ranges from several TeV down to a few hundreds of GeV, which is the prime target space for colored particle searches at LHC~\cite{Eichten:1984eu}.

This model exemplifies the approach underpinning the coannihilation codex~\cite{Baker:2015qna}, where the relic density calculation points to the region of parameter space of interest for collider searches.  Moreover, this simplified model exhibits a novel dijet resonance + missing transverse energy (MET) signature, which thus far remains an unexplored search channel at the LHC.  In contrast to ad hoc models exhibiting this collider signature~\cite{Gupta:2015lfa, Autran:2015mfa, Bai:2015nfa}, our model readily generates the resonant dijet and MET final state at colliders by recycling vertices used in the coannihilation diagram.  Other signals in direct detection and indirect detection experiments are also predicted by reordering the topology of the coannihilation diagram.  In this way, we highlight the versatility and power of the coannihilation approach where all features of the DM annihilation mechanism connect to phenomenological signatures at dark matter experiments.

In section~\ref{sec:model}, we present the field content, Lagrangian, and general phenomenology of our $s$-channel diquark mediated DM coannihilation model.  We show the relic density results for this model in section~\ref{sec:relic:density}, demonstrating that favorable regions in parameter space are within reach of the LHC.  We also briefly comment on the direct detection and indirect detection prospects.  In section~\ref{sec:existing:searches}, we review the existing collider bounds for our $s$-channel diquark mediator and our color triplet coannihilation partner, and also extrapolate these bounds to $100$ fb$^{-1}$ of 13~TeV LHC luminosity.  We study the sensitivity of our proposed dijet resonance + MET collider signature in section~\ref{sec:dq3:mixed}, highlighting the fact that this novel signature both probes new parameter space outside the reach of current searches as well as makes the DM coannihilation connection manifest.  We conclude in section~\ref{sec:conclusion}.

\section{Diquark mediated coannihilation}
\label{sec:model}

As emphasized in the coannihilation codex~\cite{Baker:2015qna},
simplified models that explicitly model the dark matter annihilation
mechanism offer unique phenomenology previously neglected in dark
matter studies.  The codex classifies dark matter simplified models by
the Standard Model (SM) gauge charges of the dark sector and mediator
fields and the $s$-channel, $t$-channel, four point, or hybrid (both
$s$-channel and $t$-channel) topology of the coannihilation diagram.
Having such a classification enables a clear and thorough exploration
of the signature space of dark matter at colliders, aiding in the
prioritization of searches and helping to identify unexplored final
states.  These experimental signatures are in fact guaranteed in this
framework, since they result from stitching together production modes
and decay vertices dictated by SM gauge interactions and the
coannihilation diagram.

\subsection{An \texorpdfstring{$s$}{s}-channel example: the ST6 model}
\label{sec:model:st6}

\begin{table}[!tb]
	\centering
	\small
	\begin{tabular}{!{\vrule width 1pt} c | c | c !{\vrule width 1pt}}
		\noalign{\hrule height 1pt}
		Field & $(SU(3)_C, SU(2)_L, U(1)_Y)$ & Spin assignment \\
		\hline
		\text{DM}  & (1, 1, 0)    & \text{Majorana fermion} \\
		\text{X}   & (3, 1, -2/3) & \text{Dirac fermion} \\
		\text{M}   & (3, 1, -2/3) & \text{Scalar} \\
		\noalign{\hrule height 1pt}
	\end{tabular}
	\caption{Field content, Standard Model gauge quantum numbers, and spin assignments for the scalar diquark case study ST6. Electric charge is defined as $Q \equiv T_3 + \dfrac{1}{2} Y$, where $T_3$ is the third component of weak isospin.}
	\label{tab:ST6fields}
\end{table}

One exemplary model to consider is ST6 (``$s$-channel mediator, color
triplet model 6'' from table 4 of reference~\cite{Baker:2015qna}), which
features an $s$-channel coannihilation topology with a fermionic color
triplet coannihilation partner X, scalar color triplet mediator M, and
a fermionic SM gauge singlet dark matter DM.  The field definitions
are shown in table~\ref{tab:ST6fields}.  This model is especially
attractive to consider because both the coannihilation partner X and
the mediator M can be pair-produced at the LHC via strong
interactions.  The corresponding large pair production rates for these
particles could lead to immediate LHC discovery prospects.  Moreover,
this model juxtaposes signatures reminiscent of supersymmetry
with unique collider signatures characteristic of $s$-channel coannihilation,
which highlights the importance of our bottom-up approach to DM model
building.  Other $s$-channel color triplet, sextet, and octet models
in the coannihilation codex~\cite{Baker:2015qna} also share much of
the same phenomenology as ST6, and thus the analysis we present is
representative of many models detailed in the codex.

Given the field content in table~\ref{tab:ST6fields}, the Lagrangian
for this model is
\begin{align} \label{eq:lagrangian:diquarktriplet}
	\mathcal{L} & = \mathcal{L}_\text{DM} + \mathcal{L}_\text{X} + \mathcal{L}_\text{M} + \mathcal{L}_\text{vis} + \mathcal{L}_\text{dark} \ , \nonumber \\
  \mathcal{L}_\mathrm{DM} & = \frac{i}{2} \overline{\mathrm{DM}} \slashed{\partial} \mathrm{DM} - \frac{m_\mathrm{DM}}{2} \overline{\mathrm{DM}} \, \mathrm{DM} \ , \nonumber \\
  \mathcal{L}_\mathrm{X} & = i \overline{\mathrm{X}} \slashed{D} \mathrm{X} - m_{\mathrm{X}} \overline{\mathrm{X}} \, \mathrm{X} \ , \nonumber \\
  \mathcal{L}_\mathrm{M} & = \left( D_\mu \mathrm{M} \right)^\dagger \left( D^\mu \mathrm{M} \right) - m_\mathrm{M}^2 \mathrm{M}^* \, \mathrm{M} - \lambda_\mathrm{HM} \left( H^\dagger H \right) \mathrm{M}^* \mathrm{M} - \lambda_\mathrm{M} \left(\mathrm{M}^* \mathrm{M} \right)^2 \ , \nonumber \\
  \mathcal{L}_\mathrm{vis} & =  - \varepsilon _{abc} \mathrm{M}_a  \overline {u_b^C } \left( {Y_{ud}^L P_L  + Y_{ud}^R P_R } \right)d_c  - Y_{u\ell } \mathrm{M}^*_a \overline {\ell _R^C } u_R^a  - Y_{QL} \mathrm{M}^*_b \left( {\overline {L_L^C } _\alpha  \varepsilon ^{\alpha \beta } (Q_L )_\beta ^b } \right) + \mathrm{h.c.} \ , \nonumber \\
  \mathcal{L}_\mathrm{dark} & = - y_D \overline{\mathrm{X}} \, \mathrm{DM} \, \mathrm{M} + \mathrm{h.c.} \ ,
\end{align}
where $H$ denotes the SM Higgs field, $a$, $b$, and $c$ denote color
indices, $\alpha$ and $\beta$ denote $SU(2)_L$ weak isospin indices,
$C$ is the charge conjugation operator, and we have suppressed flavor
indices on $Y_{ud}^L$, $Y_{ud}^R$, $Y_{u\ell}$ and $Y_{QL}$.  The
Yukawa interactions in $\mathcal{L}_\mathrm{vis}$ are written in the
mass basis for all quarks and leptons.  We need a
Cabibbo-Kobayashi-Maskawa (CKM) rotation to write $d_L$ in the mass
basis, therefore $Q_L = \left( {u_L ,V_{CKM} d_L } \right)$ for mass
eigenstates.\footnote{Note that $SU(2)_L$ invariance can be made manifest in
$\mathcal{L}_\text{vis}$ by rewriting the first Lagrangian term as
$-Y_{QQ}^{ij} \varepsilon _{abc} M_a \left( {\overline {Q_L^C }
  _\alpha ^{b,i} \varepsilon ^{\alpha \beta } (Q_L )_\beta ^{c,j} }
\right) -Y_{ud}^R \varepsilon _{abc} M_a \left( {\overline {u_R^C }
  _b d_{Rc} } \right)$, and identifying $Y_{ud}^L$ with $2 Y_{QQ}
V_{\text{CKM}}$.}

The gauge quantum numbers allow for additional couplings between X or
DM and the SM fields.  For example, X shares the same quantum numbers
as the right-handed down-type quarks, and hence a Yukawa interaction
between X, the SM Higgs, and $Q_L$ would lead to mixing between X and
the SM down-type quarks after the Higgs acquires a vev.  Then, the DM
particle would decay to three quarks via a mixing angle insertion,
which is phenomenologically ruled out.  Separately, DM could have a
Yukawa interaction with the Higgs and the left-handed lepton doublet,
which would immediately lead to mixing with active neutrinos or other
sterile neutrino phenomenology, depending on the mass generation
mechanism for SM neutrinos.  In order to ensure our DM particle is an
appropriate dark matter candidate, we impose a global $\mathbb{Z}_2$
symmetry under which both the DM and X are odd while the mediator and
all SM particles are even.  Once this $\mathbb{Z}_2$ symmetry is
introduced, only the interactions shown in
equation~\ref{eq:lagrangian:diquarktriplet} remain.

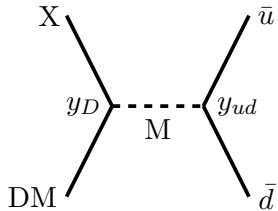
\begin{figure}[tb]
  \centering
  \begin{tikzpicture}[line width=1.4pt, scale=1.5]
	\draw[fermionna] (0.8,0.8)--(0.4,0);
	\draw[fermionna] (0.8,-0.8)--(0.4,0);
	\draw[fermionna] (-0.8,0.8)--(-0.4,0);
	\draw[fermionna] (-0.8,-0.8)--(-0.4,0);
	\draw[scalarna] (-0.4,0)--(0.4,0);
	\node at (-0.65, 0) {$y_D$};
	\node at (0.7, 0) {$y_{ud}$};
	\node at (-0.95,0.8) {X};
	\node at (-1.1,-0.8) {DM};
	\node at (0.95,0.8) {$\bar u$};
	\node at (0.95,-0.8) {$\bar d$};
	\node at (0,-0.2) {M};
\end{tikzpicture}
  \caption{Coannihilation diagram for the model ST6, with a diquark mediator M, coannihilation partner X, and dark matter DM.  The Yukawa couplings $y_D$ and $y_{ud}$ denote the mediator couplings to dark sector particles and SM particles, respectively.}
  \label{fig:coannihilationschannel}
\end{figure}

The field content in table~\ref{tab:ST6fields} and Lagrangian in
equation~\ref{eq:lagrangian:diquarktriplet} lead to the
$s$-channel coannihilation diagram depicted in
figure~\ref{fig:coannihilationschannel}.  We reiterate that the
prescription from reference~\cite{Baker:2015qna} for constructing
simplified models is based solely on the coannihilation topology and
SM gauge representation assignments.  As a result, the parameter space
of a given model may have significant constraints coming from flavor
physics, proton decay bounds, Higgs physics, and vacuum stability,
among others.  We discuss these questions in the remainder of this
section and will find that interesting and viable parameter space
regions are still allowed for model ST6, from which we motivate our
LHC study focused on dijet resonances and MET.

\subsection{Low energy constraints}
\label{sec:flavor:protondecay}

In this section, we investigate the constraints on the flavor
structure of the Yukawa matrices shown in $\mathcal{L}_\text{vis}$ in
equation~\ref{eq:lagrangian:diquarktriplet}. First, we note that the
simultaneous presence of all Yukawa matrices in
$\mathcal{L}_\text{vis}$ generates diquark-mediated diagrams that
induce proton decay, which is very strongly constrained.  For example,
if $(Y_{QL})_{11}$ and $(Y_{ud}^L)_{11} = (Y_{ud}^R)_{11} = y_{ud}$ are each
nonzero, our mediator generates the decay process $p \to e^+ \pi^0$,
which has a minimum lifetime of at least $8.2 \times 10^{33}$
years~\cite{Agashe:2014kda}.  Correspondingly, the product of these
couplings is limited to be $\sqrt{y_{ud} (Y_{QL})_{11}} < 4 \times
10^{-17} (m_{\text{M}}/ \text{GeV})$.  As usual,
however, these proton decay bounds can be avoided by extending the
accidental baryon and lepton global symmetries of the SM to the new
physics particles.  If we charge M with baryon number $B = 2/3$ and $L
= 0$, for instance, the $Y_{QL}$ and $Y_{u\ell}$ matrices identically
vanish.  We will hence turn off these couplings, enabling the dijet
resonance signature at the LHC instead of the leptoquark resonance,
given that the leptoquark case study was presented in
reference~\cite{Baker:2015qna}.

Next, we focus on possible flavor-violating entries in $Y_{ud}^L$ and
$Y_{ud}^R$, which are highly constrained by low energy flavor
violation probes.  Because our mediator has electric charge $-1/3$, it
does not induce any tree-level $\Delta F = 2$ meson mixing, which
instead occurs via box diagrams.  For instance, simultaneous
$(Y_{ud}^L)_{11}$ and $(Y_{ud}^L)_{12}$ diquark couplings induce large
$K^0$--$\bar{K}^0$ mixing, leading to the constraint $m_M \gtrsim 10^3
(Y_{ud}^L)_{11} (Y_{ud}^L)_{12}$~TeV~\cite{Bona:2007vi,
  Isidori:2010kg, Giudice:2011ak}.  If the $K^0$--$\bar{K}^0$ box
diagram involves both chiralities, the experimental constraint is more
stringent, requiring $m_M \gtrsim 10^4 (Y_{ud}^{ij})^2$~TeV where $i$,
$j$ are light quark flavor indices.  Multiple diagonal entries in
$Y_{ud}^L$ are also constrained, since nonzero $(Y_{ud}^L)_{11}$ and
$(Y_{ud}^L)_{22}$ couplings also lead to $K^0$--$\bar{K}^0$ mixing via
$M$ and $W^\pm$ mixed box diagrams.  The resulting bound requires $m_M
\gtrsim 100 \sqrt{ (Y_{ud}^L)_{11} (Y_{ud}^L)_{22}}$~TeV.  Diagonal
entries in $Y_{ud}^R$ are much less constrained by this diagram
because of a chiral quark mass suppression.

We note that while arbitrary flavor structures in
$\mathcal{L}_\text{vis}$ are incompatible with proton decay and flavor
violation constraints, these flavor aspects are orthogonal to the
primary relic density motivation of the model.  Thus, for the
remainder of the section, in addition to the $B = 2/3$ and $L = 0$
global charges of M, we will set $(Y_{ud}^L)_{11} = (Y_{ud}^R)_{11} =
y_{ud}$ and set all other entries of $Y_{ud}^L$ and $Y_{ud}^R$ to zero
to satisfy these flavor constraints.  This flavor structure, while
sufficient for our purposes in discussing the relic density connection
to LHC signatures, is overly restrictive given the constraints above,
however, and fuller discussions of possible flavor aspects of scalar
diquarks can be found in references~\cite{Isidori:2010kg,
  Giudice:2011ak}.  Finally, electroweak oblique corrections can also
constrain the X and M masses in our model, but these corrections
decouple quickly for X and M masses even moderately above the
electroweak scale~\cite{Arnold:2009ay}.

\subsection{Vacuum stability and precision Higgs physics}
\label{sec:ccbvacua:higgs}

We now consider the impact of the Higgs portal quartic coupling
$\lambda_{\text{HM}}$ in $\mathcal{L}_{\text{M}}$ for both the
stability of the electroweak vacuum and precision Higgs physics.
While this term is again irrelevant for the suitability of this
coannihilation model regarding relic density, the magnitude of this
coupling cannot be constrained by global or gauge symmetries and can
have significant phenomenological effects.

First, we note that a large, negative value for $\lambda_{\text{HM}}$
will destabilize the electroweak vacuum and cause the diquark mediator
to obtain a vacuum expectation value (vev), which is
phenomenologically unacceptable.  Hence, we require $m_\text{M}^2 +
\lambda_{\text{HM}} v^2/2 > 0$, where $v$ is the Higgs vev, $H =
\frac{1}{\sqrt{2}} (v + h)$.  In addition, we impose $\lambda_M > 0$
and $\lambda_{\text{HM}} + \sqrt{\lambda_{\text{H}}
  \lambda_{\text{M}}} > 0$, with $\lambda_{\text{H}}$ the SM Higgs
quartic coupling, to ensure a global vacuum
stability~\cite{Gunion:1989we}.

The quartic coupling $\lambda_{\text{HM}}$ is also constrained from
Higgs physics, since the mediator M propagates in the loop-induced
gluon fusion Higgs production process.  In the limit where the top
mass and $m_\text{M}$ are heavy compared to the Higgs mass, we can
calculate the gluon signal strength $\mu_{ggF} \approx \left| {1 + c_M
  \lambda_{\text{MH}} v^2 /(8m_\text{M}^2 )}
\right|^2$~\cite{Dobrescu:2011aa, Kumar:2012ww}, where $c_M = 2 \times
4/3$ is proportional to the quadratic Casimir for a color triplet
complex scalar. The ATLAS and CMS combination of Higgs measurements
constrain $\mu_{ggF}= 1.03^{+0.17}_{-0.15} $~\cite{CMS:2015kwa}, which
implies $-0.8 < \lambda _{\text{HM}} < 1.2$ for $m_\text{M} =
500$~GeV. The Higgs decay rate to photons is affected by the charged
mediator running in the loop.  However, since contributions to this
decay mode from intermediate scalars are suppressed by $v^2 /
m_\text{M}^2$~\cite{Djouadi:2005gj} and the electric charge, the
effect of the mediator on the Higgs diphoton decay width will be small
even for large $\lambda_{\text{HM}}$.

As the quartic couplings are not relevant for the collider analysis or
relic density calculation, we will simplify the remaining discussion
by neglecting the terms $\lambda_{\text{HM}}$ and
$\lambda_{\text{M}}$.  If, however, Higgs data begins to show
significant deviations from the SM expectation and a dijet signal
emerges from our proposed analysis in section~\ref{sec:dq3:mixed}, we
can revisit the impact of $\lambda_{\text{HM}}$ to both fit the
deviations to our model and predict additional signatures given by the
coannihilation topology.

\subsection{Phenomenological features}
\label{sec:pheno:widths}
As described in the previous sections, the various Yukawa structures
and quartic couplings appearing in $\mathcal{L}_\text{vis}$ and
$\mathcal{L}_\text{M}$ can lead to unacceptable proton decay, flavor
violation, vacuum instability, and Higgs physics effects.  We readily
satisfy all of these bounds, however, by giving baryon number $2/3$
and lepton number $0$ to M, restricting $(Y_{ud}^L)_{11} =
(Y_{ud}^R)_{11} = y_{ud}$ as the only nonzero entries in $Y_{ud}^L$
and $Y_{ud}^R$, and neglecting the quartic scalar couplings in
$\mathcal{L}_\text{M}$.

These simplifications lead to a small number of physical parameters in
our simplified model: the masses of the mediator, X, and DM, the
visible coupling $y_{ud}$, and the dark coupling $y_D$.  We define
$\Delta \equiv (m_\text{X} - m_\text{DM}) / m_{\text{DM}}$ to be the
fractional mass splitting between the coannihilation partner X and the
dark matter, and we always assume X is heavier than DM, which ensures
the stability of the DM particle.  Since DM is a SM gauge singlet, it
is only produced at the LHC in the decays of X or M, where both X and
M can be pair-produced from SM gauge interactions.  We also can
singly-produce M via the coupling $y_{ud}$, giving a dijet resonance
or the very difficult M $\to (jj)_\text{soft} + \slashed{E}_T$ channel.

The coannihilation partner X has only a single decay mode, X $\to$ DM
$\bar{u} \bar{d}$, which is prompt as long as $\Delta$ is not
extremely small.  Pair production of X then leads to the
$2(jj)_\text{soft} + \slashed{E}_T$ final state, where the softness of
the outgoing jets is correlated with $\Delta$.  If $\Delta$ is small,
these soft jets will escape detection and the primary signature will
be $\slashed{E}_T$ accompanied with one or two jets from initial/final
state radiation.  This signature is already probed by the ATLAS and
CMS monojet searches~\cite{Khachatryan:2014rra, Aad:2015zva} and by
the lowest multiplicity signal regions of the multijet + MET
searches~\cite{Aad:2014wea, Khachatryan:2016kdk, Atlas:2016rxq,
  Chatrchyan:2014lfa}.  For large $\Delta$ and large DM mass, the jets
from the X decay can be hard enough to pass detector thresholds,
leading to signatures akin to gluino pair production in the minimal
supersymmetric standard model (MSSM)~\cite{Alves:2011wf, Aad:2014wea}.
Moreover, because DM is a SM gauge singlet, $\Delta$ plays a dominant
role in determining the DM relic density via the X--X and X--DM
annihilation modes in the effective DM annihilation cross
section~\cite{Griest:1990kh}.

The diquark mediator can either decay via the visible coupling
$y_{ud}$ or the dark sector coupling $y_D$.  The corresponding partial
widths, assuming massless quarks, are
\begin{align}
  \Gamma \left( \mathrm{M} \to \bar{u} \bar{d} \right) = &
\frac{ y_{ud}^2 }{4 \pi} m_\mathrm{M} \ , \nonumber \\
  \Gamma \left( \mathrm{M} \to \mathrm{X} \, \mathrm{DM} \right) = &
\frac{y_D^2}{8 \pi} m_\mathrm{M} K ( \Delta, \tau_\mathrm{DM} ) \ ,
\label{eq:widths:diquarktriplet}
\end{align}
where, for on-shell M $\to$ X DM decays, $\tau_\mathrm{DM} \equiv
m_\mathrm{DM}^2 / m_\mathrm{M}^2$ and
\begin{align}
  K ( \Delta, \tau_\mathrm{DM} ) \equiv &
\left( 1 - \Delta^2 \tau_\mathrm{DM} \right)^\frac{1}{2}
\left( 1 - ( 2 + \Delta)^2 \tau_\mathrm{DM} \right)^\frac{3}{2} \ .
\label{eq:dq3:K}
\end{align}

The possible final states for M pair production are
$2(jj)_\text{res}$, $(jj)_\text{res} + (jj)_\text{soft} +
\slashed{E}_T$, and $2(jj)_\text{soft} + \slashed{E}_T$, where we
denote the decay products from X as soft because of the fractional
mass splitting $\Delta$.  While the pair production of dijet
resonances has been a traditional new physics signature, the mixed decay
$(jj)_\text{res} + (jj)_\text{soft} + \slashed{E}_T$ gives a striking
dijet resonance + MET final state which remains unexplored.
Section~\ref{sec:dq3:mixed} will present a proposed analysis and the
experimental prospects for this interesting mixed decay signature.

The relative rates between the three M~M$^*$ final states are dictated
by the visible branching fraction $B$,
\begin{align}
  B \equiv & \mathrm{Br} \left( \mathrm{M} \to \bar{u} \bar{d} \right) = \frac{y_{ud}^2}{y_{ud}^2 + \frac{1}{2} y_D^2 K ( \Delta, \tau_\mathrm{DM} )} = \frac{B_0}{B_0 + (1 - B_0) K ( \Delta, \tau_\mathrm{DM} )} \ , \nonumber \\
  B_0 \equiv & \left. \mathrm{Br} \left( \mathrm{M} \to \bar{u} \bar{d} \right) \right|_{m_{\mathrm{DM,X}} = 0} = \frac{y_{ud}^2}{y_{ud}^2 +  \frac{1}{2} y_D^2} \ ,
\label{eq:B0:definition}
\end{align}
where we introduce $B_0$ to represent the mediator M branching
fraction into SM particles at zero dark matter mass and coannihilation
partner mass.  We note that the couplings $y_{ud}$ and $y_{D}$ can
be traded for $B_0$ or $B$ and the total width of M, which
are natural variables to discuss the collider constraints and
prospects on M.  Hence, we adopt the language of $m_\text{DM}$,
$\Delta$, $m_\text{M}$, $B_0$ or $B$, and $\Gamma_\text{M}$ to tighten
the connection between the relic density calculation and collider
probes.

We implement the diquark triplet model in
\texttt{FeynRules~v2.3.18}~\cite{Christensen:2008py, Alloul:2013bka}
with the Lagrangian in equation~\ref{eq:lagrangian:diquarktriplet}
using the simplifications summarized in the beginning of this
subsection.  This implementation is publicly available in the
FeynRules model database~\cite{website:fr:database}, where an
interface with Monte Carlo generators is also available using the
\texttt{UFO}~\cite{Degrande:2011ua} format.  Our relic density
calculations in section~\ref{sec:relic:density} are performed using
\texttt{MicrOMEGAs~v4.1.8}~\cite{Belanger:2014vza} through
\texttt{CalcHEP v3.6.25}~\cite{Belyaev:2012qa} output of
\texttt{FeynRules}.  We also performed cross-checks of our relic
density calculation using
\texttt{MadDM~v2.0}~\cite{Backovic:2013dpa,Backovic:2015tpt}.  The
collider recasting discussion in section~\ref{sec:existing:searches}
and the mixed decay study shown in section~\ref{sec:dq3:mixed} use
simulated Monte Carlo events generated from
\texttt{MadGraph5~v1.5.14}~\cite{Alwall:2011uj, Alwall:2014hca}.  We
shower and hadronize the events using
\texttt{Pythia~v8.2}~\cite{Sjostrand:2007gs} and perform detector
simulation with \texttt{Delphes~v3.2}~\cite{deFavereau:2013fsa} using
the default CMS parameter card.  When indicated, matching between the
hard matrix element and the parton shower is performed using the MLM
matching scheme~\cite{Mangano:2002ea}.

\section{Relic density}
\label{sec:relic:density}

The central concern for the simplified models developed in the coannihilation codex~\cite{Baker:2015qna} is the dark matter relic density.  By modeling the dark matter annihilation mechanism, we can directly connect the parameter space of a given simplified model to phenomenological signals in direct detection experiments, indirect detection experiments, and colliders.  Hence, the relic density calculation features strongly in motivating our discussion of experimental signatures, and we can demonstrate the particular interplay between different terrestrial and astrophysical probes manifest in our diquark mediator case study.

In this vein, we calculate the parameter space that gives the correct dark matter relic density, $\Omega h^2 = 0.1198 \pm 0.0026$~\cite{Ade:2015xua, Agashe:2014kda} for the ST6 model, subject to the Lagrangian flavor restrictions discussed in section~\ref{sec:model}.  The three fields, DM, X, and M, and two couplings, $y_{ud}$ and $y_D$, lead to several channels that determine the final DM relic density.  Since the dark matter candidate is a pure SM gauge singlet fermion, the usual pair annihilation channel $\overline{\text{DM}}$ DM $\to$ $\overline{\text{SM}}$ SM vanishes at tree level and can be neglected.  Obtaining the appropriate relic density for DM therefore depends crucially on X and M.  In particular, DM and X readily stay in thermal equilibrium with SM particles until thermal freeze out from Hubble expansion in the early universe, where the process that freezes out last determines the relic density~\cite{Griest:1990kh}.

There are four categories of two-to-two (co)annihilation channels relevant for the diquark model:
\begin{description}
	\item[\textbf{DM X $\to$ SM$_1$ SM$_2$:}] the standard coannihilation channel where DM and X coannihilate to SM particles $\bar{u} \bar{d}$ through the $s$-channel diquark mediator.  This channel is dominant only when the diagram is resonant, that is when $m_\mathrm{DM} + m_\mathrm{X} \sim m_\mathrm{M}$.
	\item[\textbf{$\overline{\text{X}}$ X $\to$ $\overline{\text{SM}}$ SM:}] pair annihilation of X into SM pairs.  Since X is a color triplet, pair annihilation of X into gluons and quarks is typically dominant, especially for low X masses when other channels are kinematically closed.  As long as DM and X are in chemical equilibrium, this channel sets a lower bound on $m_\mathrm{DM}$ for a given value of $\Delta$: ever lighter DM and X masses cause $\Omega h^2$ to fall below the measured DM relic density and hence underclose the universe.
	\item[\textbf{DM X $\to$ M $V_\mathrm{SM}$:}] real radiation of a SM gauge boson ($V_\mathrm{SM} = g$, $\gamma$, $Z$) from the on-shell production of M.  This category of channels opens up whenever $m_\mathrm{DM} + m_\mathrm{X} \gtrsim m_\mathrm{M}$ and is generated whenever X or M radiates off a SM gauge boson.  Since X and M are colored, this category is dominated by the gluon radiation diagram.
	\item[\textbf{$\overline{\text{DM}}$ DM $\to$ M$^*$ M:}] DM pair annihilation to a pair of mediators.  The last channel only opens up when $m_\mathrm{DM} \gtrsim m_\mathrm{M}$. This pair annihilation mode of DM is generated through a $t$-channel X exchange and gives the only contribution to the effective coannihilation cross section which is not exponentially suppressed by $\Delta$. The $\overline{\text{X}}$ X $\to$ M$^*$ M process also exists but is subdominant to X pair annihilation to SM particles.
\end{description}

We note that the Sommerfeld effect modifies the cross section for the process $\overline{\text{X}}$ X $\to \bar{q} q$, $gg$, where the $t$-channel exchange of soft gluons can potentially lead to significant non-perturbative effects. We use the calculation of the Sommerfeld enhancement factors for colored particles in reference~\cite{deSimone:2014pda} and implement this in \texttt{MicrOMEGAs}~\cite{Ibarra:2015nca}. In contrast to the results of reference~\cite{deSimone:2014pda}, however, we find that the Sommerfeld enhancement effects are less than a few percent on the relic density for the parameters we consider because of a cancellation between the enhancement of the cross section in the channel $\overline{\text{X}}$ X $\to gg$ and the suppression in the channel $\overline{\text{X}}$ X $\to \bar{q} q$.  We therefore neglect this effect in our relic density calculations.

\begin{figure}[tb]
  \centering
  \includegraphics[width=0.8\textwidth]{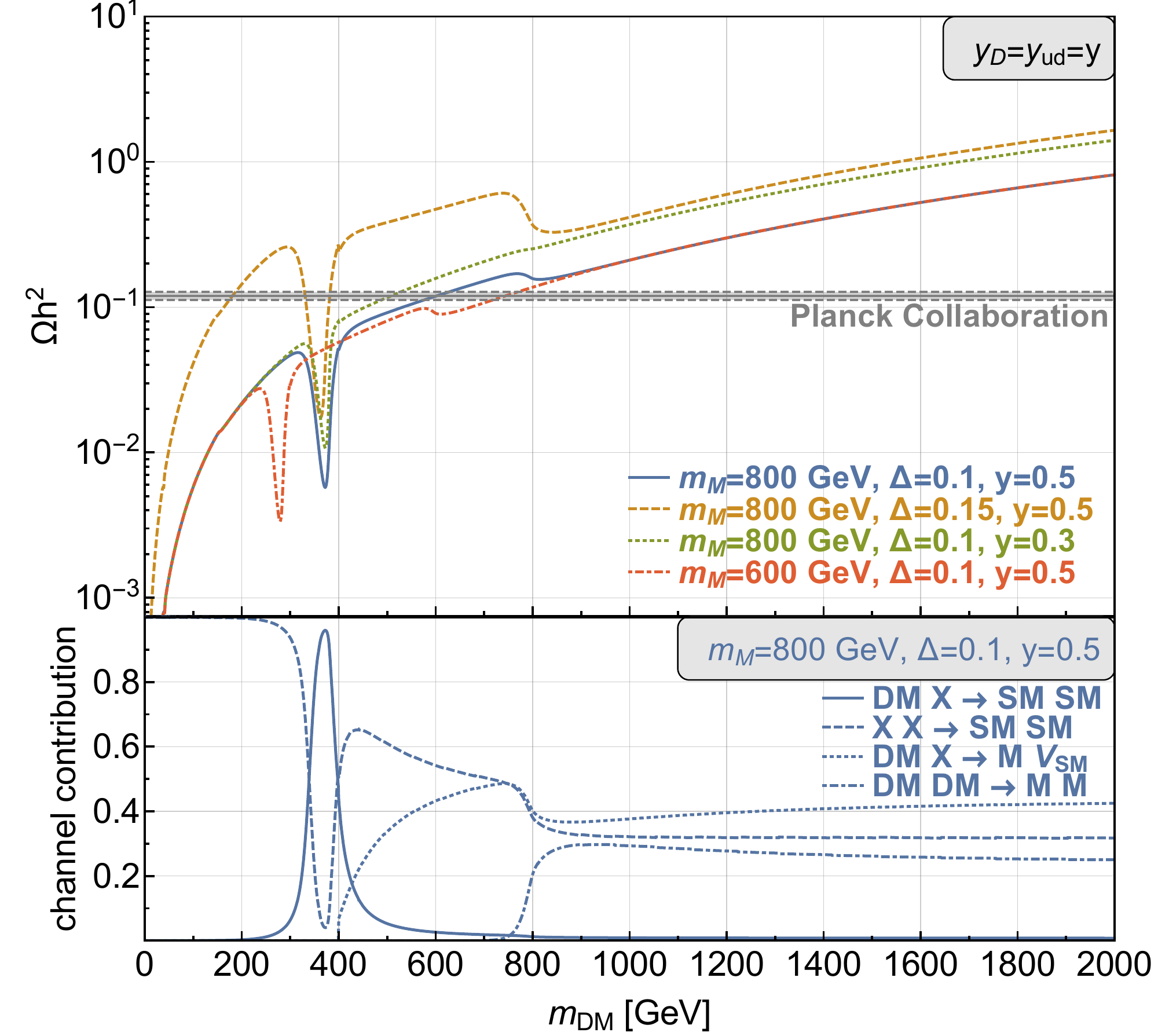}
  \caption{Upper panel: the relic density in the triplet diquark model ST6 as a function of the dark matter mass for given values of $m_\text{M}$, fractional mass splitting $\Delta$, and $y_{ud} = y_D = y$, where the gray horizontal band shows the Planck measurement with $3\sigma$ uncertainty~\cite{Ade:2015xua}. Lower panel: the relative contributions from the four different channels described in the text are shown for a single benchmark point.}
  \label{fig:dq3:relic:density:mdm}
\end{figure}

In figure~\ref{fig:dq3:relic:density:mdm}, we plot the dark matter relic density as a function of the DM mass for various choices of mediator mass, $\Delta$, and Yukawa couplings $y_{ud} = y_D = y$.  We also overlay the measured relic density with $3\sigma$ uncertainties from Planck~\cite{Ade:2015xua} as a gray horizontal band.  In the bottom panel, the relative contributions from the four different annihilation categories are displayed for a single benchmark point. This breakdown of the overall relic density calculation into constituent annihilation and coannihilation channels reinforces the importance of explicitly accounting for all fields and diagrams that contribute to the dark matter relic density.

\subsection{Relic density favored regions}
\label{sec:relic:density:regions}
We now use the relic density calculation to motivate masses and couplings to probe with collider experiments, while direct detection and indirect detection prospects are discussed in section~\ref{sec:dd:id}.  From figure~\ref{fig:dq3:relic:density:mdm}, we see that perturbative Yukawa couplings for TeV-scale mediators and electroweak scale dark sector particles generally give the correct DM relic density.  As a result, because M and X can be pair produced via strong interactions at the LHC, we expect a dynamic interplay between the parameter space region favored by the relic density calculation, the current exclusion bounds from ATLAS and CMS, and the discovery prospects from future searches.  Following section~\ref{sec:model}, we adopt $m_\text{DM}$, $\Delta$, $m_\text{M}$, $B_0$ and $\Gamma_\text{M}$ as the five parameters of our model.  Recall $B_0$ from equation~\ref{eq:B0:definition} is the particular coupling relation defining the zero mass visible branching fraction of the diquark mediator, and $\Gamma_\text{M} = \Gamma \left( \text{M} \to \bar{u} \bar{d} \right) + \Gamma \left( \text{M} \to \text{DM} \text{X} \right)$ from equation~\ref{eq:widths:diquarktriplet} is the total mediator width.

\begin{figure}[tb]
	\centering
	\includegraphics[width=0.8\textwidth]{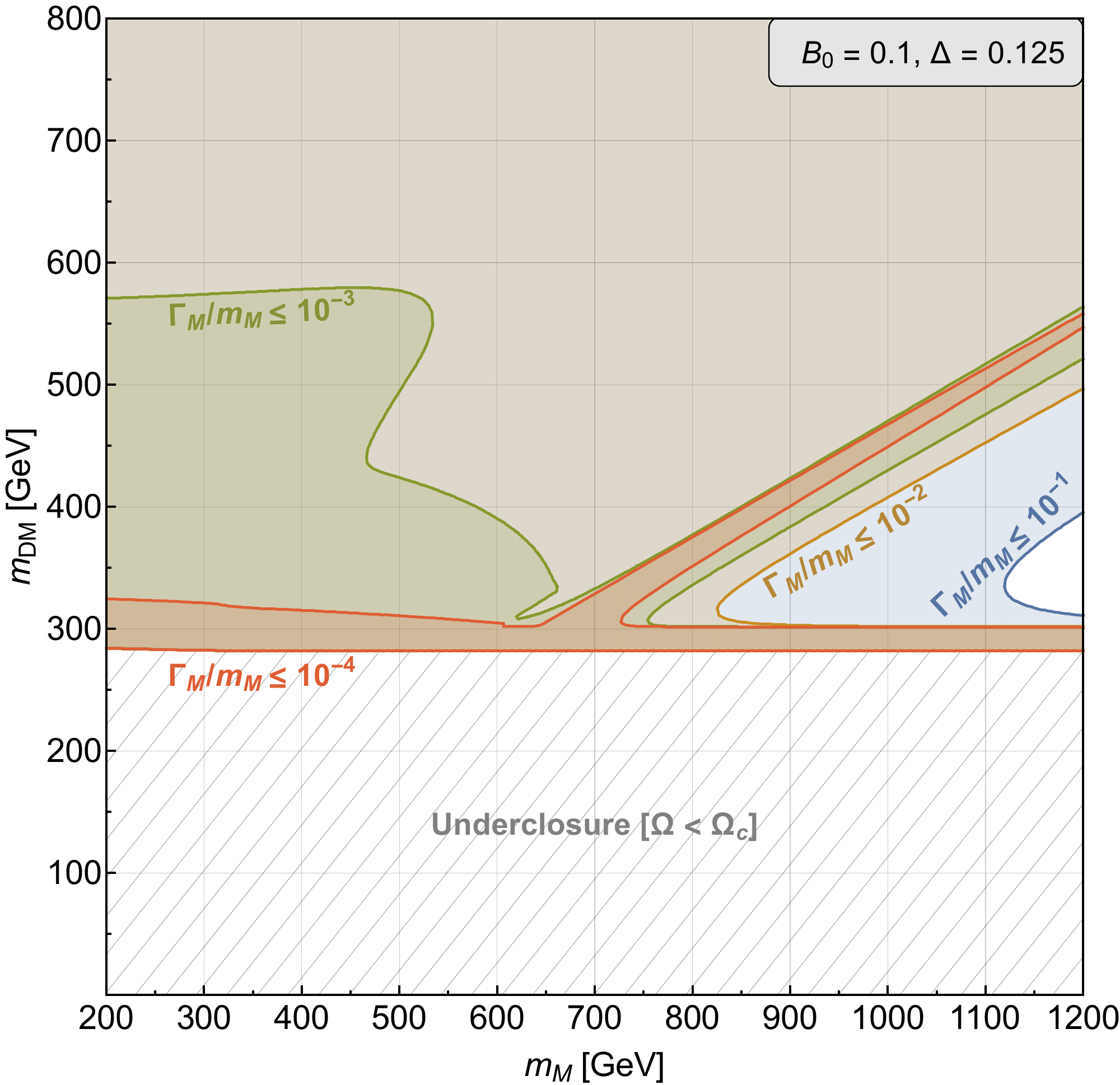}
	\caption{Parameter space consistent with the measured relic density within $3\sigma$ from Planck~\cite{Ade:2015xua}, for fixed $B_0 = 0.1$ and $\Delta = 0.125$ in the $m_\text{DM}$ versus $m_\text{M}$ plane, with different overlays of maximal $\Gamma_\text{M} / m_{\text{M}}$.}
	\label{fig:dq3:relic:density:Gamma}
\end{figure}

In figure~\ref{fig:dq3:relic:density:Gamma}, we scan the $m_\text{DM}$ vs.~$m_\text{M}$ plane for $B_0 = 0.1$ and $\Delta = 0.125$, where $B_0 = 0.1$ is equivalent to requiring $y_D^2 = 18 y_{ud}^2$.  We overlay regions with $\Gamma_\text{M} / m_\text{M} \leq 10^{-1}$, $10^{-2}$, $10^{-3}$, and $10^{-4}$, which effectively scan over $y_{ud}$ and $y_D$ given the proportionality above.  We see that heavy DM masses, which generally overclose the universe as depicted from the asymptotic behavior in figure~\ref{fig:dq3:relic:density:mdm}, still recover the correct $\Omega h^2$ if the DM X $\to$ SM$_1$ SM$_2$, DM X $\to$ M $V_{\text{SM}}$, and $\overline{\text{DM}}$ DM $\to$ M$^*$ M channels are large enough.  If the width of the mediator is small, however, the DM X $\to$ M $V_\text{SM}$ and $\overline{\text{DM}}$ DM $\to$ M$^*$ M channels become negligible.  Then, the correct relic density is only found in the resonant coannihilation region, or when the everpresent $\overline{\text{X}}$ X $\to$ $\overline{\text{SM}}$ SM channel together with the given $\Delta$ produce the appropriate effective DM annihilation cross section to fall into the measured band of $\Omega h^2$.  The upper and lower lobes for $\Gamma_\text{M} / m_\text{M} \leq 10^{-3}$ show the separate impact of the $\overline{\text{DM}}$ DM $\to$ M$^*$ M and DM X $\to$ M $V_\text{SM}$ channels, respectively, avoiding overclosure of the universe by supplementing the $\overline{\text{X}}$ X $\to$ $\overline{\text{SM}}$ SM channel.  Because our model always includes the $\overline{\text{X}}$ X $\to$ $\overline{\text{SM}}$ SM channel, and because we always assume a sufficient $y_D$ coupling to keep DM and X in chemical equilibrium~\cite{Baker:2015qna}, we underclose the universe for $\mathcal{O}(100 \text{ GeV})$ DM masses for $\Delta = 0.125$, regardless of $B_0$.  For these light DM masses, the SM gauge interactions ensure that the $\overline{\text{X}}$ X $\to$ $\overline{\text{SM}}$ SM process rapidly depletes the DM relic density.  To have a viable DM model in this underclosure region, we would either have to increase $\Delta$, which expressly suppresses the $\overline{\text{X}}$ X $\to$ $\overline{\text{SM}}$ SM contribution to the effective DM annihilation cross section, or we would require additional field content outside of our simplified model.

\begin{figure}[tb]
	\centering
	\includegraphics[width=0.8\textwidth]{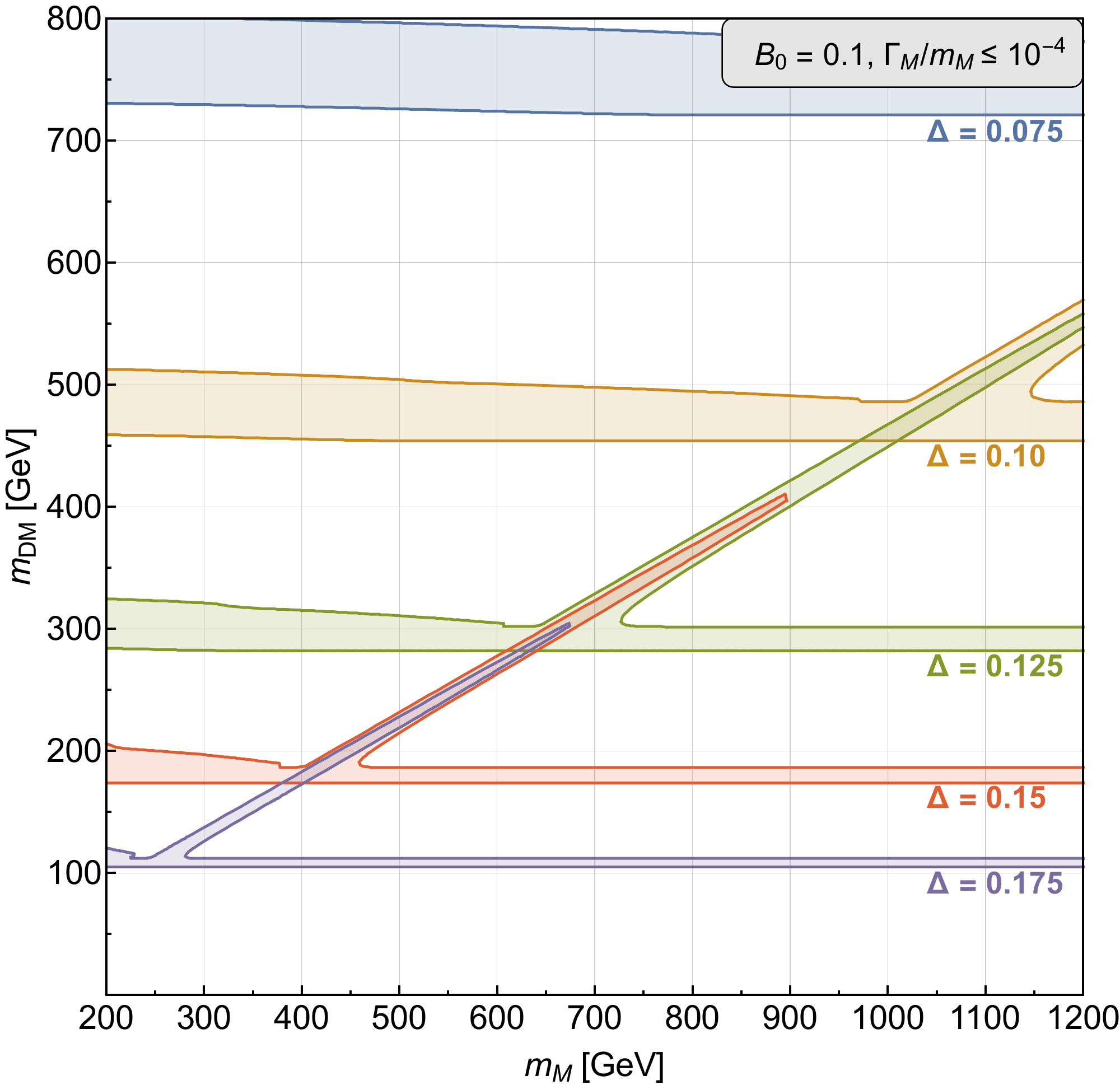}
	\caption{Parameter space consistent with the measured relic density within $3\sigma$ from Planck~\cite{Ade:2015xua}, for fixed $B_0 = 0.1$ and scanning over $\Gamma_\text{M} / m_\text{M} \leq 10^{-4}$, with different choices of $\Delta$ overlaid.}
	\label{fig:dq3:relic:density:Delta}
\end{figure}

This interesting dependence on $\Delta$ is shown in figure~\ref{fig:dq3:relic:density:Delta}, where we have chosen $B_0 = 0.1$ and scanned the Yukawa couplings such that $\Gamma_\text{M} / m_\text{M} \leq 10^{-4}$.  Most strikingly, we see that the choice of $\Delta$ sensitively changes the boundary between underclosure and overclosure delineated by the $\overline{\text{X}}$ X $\to$ $\overline{\text{SM}}$ SM process, shown in the horizontal shaded bands.  The diagonal bands aligned with $m_{\text{DM}} + m_{\text{X}} \sim m_\text{M}$ correspond to the resonant coannihilation process.  We see that the horizontal bands become wider for larger $m_{\text{DM}}$, and this growth is larger on the left side of the coannihilation funnel compared to the right side. The overall width of the bands is determined by the relative flattening of the relic density curve as $m_{\text{DM}}$ increases, as evident in figure~\ref{fig:dq3:relic:density:mdm}.  On the left side of the coannihilation funnel, however, the opening of the DM X $\to$ M $V_\mathrm{SM}$ and $\overline{\text{DM}}$ DM $\to$ M$^*$ M channels enhance the effective annihilation cross section and avoid DM overclosure of the universe.

We see from figure~\ref{fig:dq3:relic:density:Delta} that the choice of $\Delta$ effectively sets the dark matter mass scale at which the correct relic abundance is achieved.  On the other hand, $m_\text{DM}$ and $\Delta$ directly set the energy scale of the X decay products in the X $\to \bar{u} \bar{d}$ DM decay and hence control the complementarity between searches for X pair production and probes for the diquark mediator. Intriguingly, the combination of $m_{\text{DM}}$ from $300$~GeV to $500$~GeV and $\Delta \sim 0.1$ results in a spectrum of X decay products that straddles the jet thresholds used in current multijets and MET searches~\cite{Aad:2014wea}.  Therefore, we present a detailed discussion of the current collider bounds for signatures of both X and M in section~\ref{sec:existing:searches}.

Finally, we briefly comment on generalizing these conclusions to the entire set of codex models with a strongly interacting coannihilation partner X and a SM gauge singlet DM.  Given X and DM are in chemical equilibrium, the $\overline{\text{X}}$ X $\to$ $\overline{\text{SM}}$ SM annihilation channel will always dominate for low DM masses and hence connect the choice of $\Delta$ to a particular relic density motivated energy scale for $m_\mathrm{DM}$, leading to the behavior seen in figure~\ref{fig:dq3:relic:density:Delta}.  In general, larger QCD representations for X will increase the efficacy of this coannihilation channel, driving the relic density motivated scale of $m_\mathrm{DM}$ even lower for a given $\Delta$.  Moreover, the different X and DM representations will change the relative importance of the non-perturbative Sommerfeld effect~\cite{deSimone:2014pda,Ibarra:2015nca}.  A full quantitative description of relic density constraints for dark matter models with a strongly interacting coannihilation partner will be the subject of a future work.

\subsection{Direct and indirect detection prospects}
\label{sec:dd:id}
We now address the direct detection prospects for the ST6 $s$-channel diquark mediator DM model.  For $\Delta \sim 0.1$ and X and DM masses in the few hundred GeV range, the coannihilation partner X decays promptly and will have vanishing relic density today.  Since DM has no SM gauge quantum numbers, all of the DM direct detection interactions are loop-induced and scale at least with the new physics coupling $y_D$.  These loop-induced diagrams, though, are not inherently responsible for determining the dark matter relic density and thus the relic density requirement will not have a meaningful bearing on the direct detection prospects.

Since the momentum transfer in DM--nucleon scattering is $\mathcal{O}(100~\mathrm{MeV})$, and our DM mass scale is at least 100~GeV, we can use an effective field theory description based on the Lagrangian in equation~\ref{eq:lagrangian:diquarktriplet} and integrating out M, X, and the high momentum modes of DM to analyze the leading DM direct detection prospects.  We find three loop-induced operators that can drive DM direct detection:
\begin{description}
\item [{$\overline{\text{DM}}$ DM $H^\dagger H$:}] This is the leading dimension-five operator, which is loop-induced by X and M internal legs and requires a nonzero Higgs portal coupling $\lambda_{\text{HM}}$.  This operator scales as $(y_D^2 \lambda_{\text{HM}})/(16 \pi^2 m_\text{M})$ and effects DM direct detection once the Higgs acquires a vev.  Since the quartic coupling $\lambda_{\text{HM}}$ is arbitrary, however, the impact of this operator for direct detection can be completely negligible.
\item[{$\overline{\text{DM}}$ DM $\bar{u}u$, $\overline{\text{DM}}$ DM $\bar{d} d$:}] These dimension-six operators arise from insertions of X, M and quark fields and can be written schematically as $\left[ \overline{\text{DM}} \, \Gamma^\text{DM} \, \text{DM} \right] \times \left[ \bar{q} \, \Gamma^q \, q \right]$, where $\Gamma^\text{DM}$ and $\Gamma^q$ enumerate the possible Lorentz-invariant contractions (see, e.g. references~\cite{Goodman:2010yf, Goodman:2010ku}).  Because the DM is a Majorana particle, however, the pure vector $\Gamma^\text{DM} = \gamma^\mu$ and tensor $\Gamma^\text{DM} = \sigma^{\mu \nu}$ contractions vanish.  The axial vector current $\Gamma^\text{DM} = \gamma^\mu \gamma^5$ is also absent, because the hypercharge couplings of X and M induce purely vector $Z$ and photon currents after electroweak symmetry breaking.  The remaining possibilities are scalar bilinears, which require a chiral flip in the quark sector and are thus suppressed by $m_q/m_\text{M}^3$, which is too small to be constrained by direct detection.
\item[{$\overline{\text{DM}}$ DM $G_{\mu \nu}^a G^{a, \mu \nu}$:}] This dimension-seven operator is induced by X and M internal legs. After integrating out the internal legs, this operator scales with $(\alpha_s y_D^2) / (4 \pi m_\text{M}^3)$, and the strong suppression by the loop factor and three powers of $m_M$ implies we can safely ignore its contribution to the DM direct detection cross section.  Similarly, the dimension-seven operator with $B_{\mu \nu} B^{\mu \nu}$ instead of $G_{\mu \nu}^a G^{a, \mu \nu}$ is further suppressed by $\alpha' / \alpha_s$.
\end{description}
In summary, the direct detection interactions in our model are all driven by higher dimensional operators, and if $\lambda_\text{HM}$ is small, none of these operators are expected to give a meaningful direct detection DM--nucleon cross section.

Regarding indirect detection signals, we can have loop-induced pair annihilation of DM using the operators listed above.  The first operator results in final state Higgs bosons or any pair of SM gauge bosons or matter particles, while the second and third operators give final state quarks and gluons, and the aforementioned $\overline{\text{DM}}$ DM $B_{\mu \nu} B^{\mu \nu}$ dimension-seven operator gives photons and $Z$ bosons.  If $m_\mathrm{DM} < m_\mathrm{M}$, then DM can annihilate to $u d \bar{u} \bar{d}$ via one or two off-shell diquark mediators, but this process is phase-space suppressed compared to the dimension-six scalar bilinear operator.  The operator scaling behavior from the direct detection discussion also applies for the indirect detection signals, given the lack of resonant enhancement in these processes.  If $m_\mathrm{DM} > m_\mathrm{M}$, we have the pair annihilation of DM into M$^*$ M, which then decay to quarks. Because DM is a Majorana fermion, however, this annihilation channel is $p$-wave suppressed.  Therefore, the indirect detection signals from this model are dim.

\section{Existing searches}
\label{sec:existing:searches}
In this section, we discuss the existing collider constraints on our triplet diquark model. The visible mediator decays are constrained by searches for single and pair-produced dijet resonances, while the coannihilation partner X and the DM are probed via monojet searches as well as multijets + MET searches. These collider signatures arise from stitching together gauge interactions of M and X with the interaction vertices inherent in the coannihilation diagram in figure~\ref{fig:coannihilationschannel}.  

Following the discussion in section~\ref{sec:pheno:widths}, we recast the searches for single dijet resonances, pair-produced dijet resonances, monojets, and MSSM gluinos to constrain the masses and couplings of our diquark mediator, coannihilation partner, and the DM. A striking signature absent in current searches by ATLAS and CMS, however, is the promising mixed decay channel from pair-produced mediators, which we present in section~\ref{sec:dq3:mixed}.

\subsection{Single and paired dijet resonance searches}
\label{sec:dq3:dijet}
Dijet resonance searches set bounds on the mediator mass and couplings in this model.  Our relic density results in figures~\ref{fig:dq3:relic:density:Gamma} and~\ref{fig:dq3:relic:density:Delta} point to mediator masses of 200~GeV to 1200~GeV as especially attractive to target via dijet resonance searches.  Furthermore, since our diquark mediator is a color triplet, single dijet searches and paired dijet searches probe complementary sets of new physics couplings.  

Many dijet resonance searches constrain the $p p \to \mathrm{M} \to \bar{u} \bar{d}$ channel (for an overview, see reference~\cite{Dobrescu:2013coa}, and see reference~\cite{Chala:2015ama} regarding the DM context). The latest $8$~TeV search from CMS~\cite{Khachatryan:2016ecr} uses data scouting to constrain dijet resonances as light as $500$~GeV, and the $8$~TeV ATLAS search~\cite{Aad:2014aqa} uses pre-scaled jet triggers to obtain limits on resonances as light as $250$~GeV.  The $ud$ diquark nature of our mediator dictates stronger single resonance constraints from $pp$ colliders instead of $p \bar{p}$ colliders like Sp$\overline{\mathrm{p}}$S and Tevatron, hence we do not present the UA2~\cite{Alitti:1993pn}, CDF Run~1~\cite{Abe:1997hm}, or CDF Run~2~\cite{Aaltonen:2008dn} constraints.

In the CMS scouting analysis, events are required to have $H_T > 250$~GeV, where $H_T$ is the scalar $p_T$ sum of jets with $p_T > 30$~GeV and $|\eta| < 2.5$ and jets are clustered using the anti-$k_T$ algorithm with distance parameter $R = 0.5$.  The leading and subleading jets in $p_T$ are used as seeds to form wide jets, which add together jets closer than $R = 1.1$ to the nearer seed jet.  The wide jets must have $\Delta \eta < 1.3$ to reduce QCD background and an invariant mass larger than 390~GeV to ensure a smoothly falling spectrum.  The ATLAS search clusters jets with the anti-$k_T$ algorithm and $R = 0.6$, and events must have two jets with $p_T > 50$~GeV and rapidity $|y| < 2.8$.  These jets must have a rapidity separation $\Delta y < 1.2$ and an invariant mass $m_{jj} > 250$~GeV.

\begin{figure}[tb]
  \centering
  \includegraphics[width=0.8\textwidth]{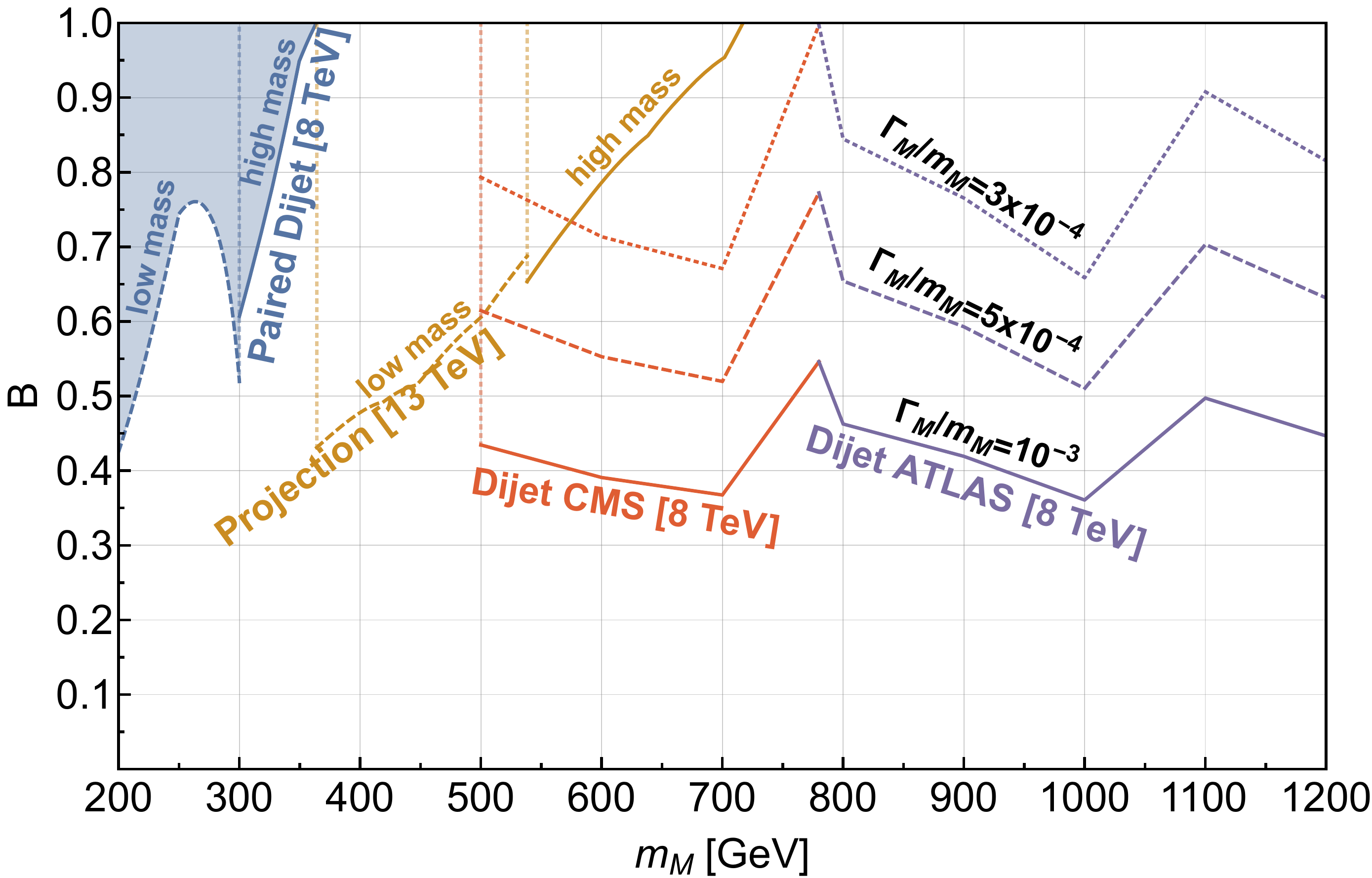}
  \caption{The existing limits from searches for paired dijet resonances are shown, where the CMS search (blue line)~\cite{Khachatryan:2014lpa} is most constraining. The orange line shows the projected exclusion by the same search at $13$ TeV and $100$ fb$^{-1}$ of integrated luminosity.  We also show the envelope of dijet resonance searches by CMS (red)~\cite{Khachatryan:2016ecr} and ATLAS (purple)~\cite{Aad:2014aqa} as exclusion contours for different choices of $\Gamma_\mathrm{M} / m_\mathrm{M} = 3 \times 10^{-4}$ (dotted), $5 \times 10^{-4}$ (dashed), and $10^{-3}$ (solid). Only the most constraining dijet resonance search is displayed for a given mediator mass.}
  \label{fig:exclusion:dijet}
\end{figure}

Given a narrow mediator M, with $\Gamma_\text{M} / m_\text{M} \lesssim 10^{-1}$, the single dijet production rate scales as $\sigma (y_0, m_{\text{M}}) (y_{ud}^2 / y_0^2) B$, where $\sigma (y_0, m_\text{M})$ is the cross section calculated using a reference visible coupling $y_0$ and mediator mass $m_\text{M}$ and $B$ is the visible branching fraction as defined in equation~\ref{eq:B0:definition}.  For diquarks decaying purely to dijets, $B = 100\%$ and the bound on the coupling is extracted by equating $\sigma(y_0, m_{\text{M}}) (y_{ud, \text{excl}}^2 / y_0^2) = \sigma_{\text{excl}}$ for an excluded cross section $\sigma_{\text{excl}}$ from experiment.  When the X DM decay is kinematically accessible, though, it is more convenient to constrain the visible branching fraction $B$ for fixed choices of the total width $\Gamma_\text{M}$, since these parameters encode the entire dependence of the signal dijet rate.  The excluded visible branching fraction is determined by equating $\sigma (y_0, m_{\text{M}}) (y_{ud}^2 / y_0^2) B_\text{excl} = \sigma (y_0, m_{\text{M}}) B_\text{excl}^2 \Gamma_\text{M} / \Gamma_0 = \sigma_{\text{excl}}$, where $\Gamma_0 = y_0^2 \text{M} / (4 \pi)$ is the visible width determined by the reference coupling $y_0$.  These exclusions from CMS~\cite{Khachatryan:2016ecr} and ATLAS~\cite{Aad:2014aqa} are shown in figure~\ref{fig:exclusion:dijet} for $\Gamma_\text{M} / m_\text{M} = 3 \times 10^{-4}$, $5 \times 10^{-4}$, and $10^{-3}$.  We see that larger total mediator widths lead to stronger exclusions, since the overall dijet resonance rate increases with the total mediator width.  Nevertheless, single dijet constraints are not sensitive to total widths smaller than $\Gamma_\text{M} / m_\text{M} = 10^{-4}$, hence we use $\Gamma_\text{M} / m_\text{M} = 10^{-4}$ as an upper limit on the total mediator width in our calculations in section~\ref{sec:dq3:mixed}.

Pair-produced diquarks are dominantly produced via color interactions, and the corresponding paired dijet signal scales with $B^2$ if the dark decay M $\to$ X DM is kinematically open.  For $m_\text{M} < m_\text{DM} + m_\text{X}$, though, the dark decay is closed and paired dijet searches probe mediator masses independently of the visible coupling $y_{ud}$.  The single and paired dijet searches are otherwise complementary, given that the single dijet mass reach is typically higher than the paired dijet mass reach.

There are paired dijet searches by CMS using $7$~TeV~\cite{Chatrchyan:2013izb} and $8$~TeV~\cite{Khachatryan:2014lpa} data.  Each dijet pair must fulfill requirements on $\Delta R_{jj}$ and then the combination of the two pairs that minimizes $\Delta m / m_\mathrm{avg}$ is selected.  To suppress the QCD continuum background, a further cut on the $p_T$ imbalance versus the average mass of the dijet pairs is imposed, ensuring a smoothly falling distribution in the average mass. This distribution is used to obtain cross section limits on a pair-produced dijet resonance with masses between $200$~GeV and $1000$~GeV for the $8$~TeV search, which supersedes the old result.  There is also an ATLAS $7$~TeV analysis~\cite{ATLAS:2012ds}, but this covers the mass window ranging from $150$~GeV to $350$~GeV and is less constraining than the $8$~TeV search by CMS.

Hence, we use the most constraining $8$~TeV CMS analysis~\cite{Khachatryan:2014lpa}, which gives the cross section limit as $\sigma \times B^2$. We recast their limit using \texttt{NLL-fast~v2.1}~\cite{Borschensky:2016xy, Beenakker:2009ha, Kulesza:2009kq, Kulesza:2008jb} in combination with \texttt{CTEQ~v6.6}~\cite{Nadolsky:2008zw} parton distribution functions (PDFs) to calculate $m_\mathrm{M}$ dependent next-to-leading order (NLO) cross sections at $8$~TeV LHC. The resulting limits are shown in figure~\ref{fig:exclusion:dijet} as the shaded blue region, where we have delineated the disjoint low mass and high mass search regions. We see that even low mass mediators are unconstrained by the data if the visible branching fraction is less than 40\%.

Using \texttt{Collider~Reach}~\cite{Salam:2014xy}, we estimate the 13~TeV sensitivity for pair produced dijet resonance searches, assuming an integrated luminosity of $100$~fb$^{-1}$.  A similar projection for the dijet resonance limits is not given because these would only alter the maximum value for $\Gamma_\mathrm{M} / m_\mathrm{M}$, and because the mass sensitivity using data scouting techniques by CMS or prescaled jet triggers by ATLAS has not been demonstrated at 13~TeV.  Both the low mass and high mass paired dijet search regions are scaled upward, although this is an oversimplification of the potential search strategy at 13~TeV LHC. We expect these rough limits are nonetheless indicative of the possible sensitivity, given that \texttt{Collider~Reach} simply uses PDF scalings to mimic the production of on-shell resonances when the background and signal are generated by the same partonic scattering channel. The projection is shown as an orange contour in figure~\ref{fig:exclusion:dijet}, where again, visible branching fractions below 40\% are not expected to be constrained.

\subsection{Jets + MET searches}
\label{sec:dq3:jets:met}
In contrast to the resonance signatures from our diquark mediator, the coannihilation partner X decays to soft jets and missing transverse energy with a 100\% branching fraction.  Pair production of X can then be constrained by monojet searches as well as multijet + MET searches. The resolved nature of the soft jets depends on the fractional mass splitting $\Delta$ and the X mass, which influences the tradeoff in signal sensitivity between the monojet and multijet + MET searches.  Jets from initial state radiation (ISR), however, skew the search sensitivity to the multijet + MET searches, especially for light X masses.

Since X is a color triplet fermion, we can calculate the pair production cross section for X using \texttt{Top++~v2.0}~\cite{Czakon:2011xx} interfaced with \texttt{LHAPDF~v6.1}~\cite{Buckley:2014ana} and the \texttt{NNPDF~v3.0}~\cite{Ball:2014uwa} set to obtain NLO signal cross sections for arbitrary X masses.  Using \texttt{CheckMate~v1.2}~\cite{Drees:2013wra}, we find the 8~TeV jets + $\slashed{E}_T$ search by ATLAS~\cite{Aad:2014wea} imposes the strongest constraints on the X mass for various choices of $\Delta$.  This search considers a wide range of signal regions characterized by the number of hard jets observed in the event and the value of the effective mass $m_\mathrm{eff}$, defined as the scalar $p_T$ sum of all jets with $p_T > 40$~GeV plus the $\slashed{E}_T$. The most powerful region for our signal is the ``2-jet medium'' region, that, in particular, requires $m_\mathrm{eff} > 1.2$~TeV in addition of two hard jets. Since $\Delta$ is typically small, the two hard jets observed for our signature come from ISR.  The relatively relaxed event selection implies this search region is systematics limited, where ATLAS quotes a $6.6\%$ systematic uncertainty for a background expectation of $760$ events.  The related ``2-jet tight'' and ``4-jet loose'' regions, with $m_\text{eff}$ cuts of $1.6$ and $1.0$~TeV, respectively, are also sensitive to our final state but are less powerful than the ``2-jet medium'' region and have larger systematic uncertainties of about 8\%.  We remark that although the strongest bounds for this search are obtained by tagging on ISR jets, the traditional ATLAS monojet search~\cite{Aad:2015zva} leads to much weaker bounds for our choices of $\Delta$.  On the other hand, we expect that dedicated analyses beyond the monojet and multijet + MET searches can improve the sensitivity to compressed spectra~\cite{Nath:2016kfp, Delgado:2016gqn}.

We show the 95\% confidence level (C.L.) exclusion limits on the X mass for $\Delta = 0.1$, $0.125$, and $0.15$ in table~\ref{tab:limits:jetsplusmet}.  Note we neglect the impact of the mediator pair production here, which can give the same final state via the M$^*$ M $\to 2(jj)_\mathrm{soft} \slashed{E}_T$ decay chain.  Since this decay is only kinematically open once $m_\mathrm{M} \geq m_\mathrm{X} + m_\mathrm{DM}$, this process pays the suppression factors of a much smaller pair production cross section of M and the dark decay branching fraction $(1-B)^2$ compared to the X pair production rate.  Our results rule out X masses below about 390~GeV for the $\Delta \sim 0.1$ values motivated by figure~\ref{fig:dq3:relic:density:Delta}.

\begin{table}
	\centering
	\begin{tabular}{!{\vrule width 1pt} c !{\vrule width 1pt} c | c | c !{\vrule width 1pt}}
		\noalign{\hrule height 1pt}
		Search & $\Delta = 0.1$ & $\Delta = 0.125$ & $\Delta = 0.15$ \\
		\noalign{\hrule height 1pt}
		$8$~TeV & $384$~GeV & $396$~GeV & $392$~GeV \\
		$13$~TeV & $398$~GeV & $399$~GeV & $396$~GeV \\
		$13$~TeV, $100$~fb$^{-1}$ projection & $464$~GeV & $468$~GeV & $477$~GeV \\
		\noalign{\hrule height 1pt}
	\end{tabular}
	\caption{The 95\% C.L.~exclusion limits on $m_\mathrm{X}$ from $8$~TeV and $13$~TeV jets + MET searches by ATLAS~\cite{Aad:2014wea, Atlas:2016rxq} at the LHC, as well the projected exclusion reach for $100$~fb$^{-1}$
	of $13$~TeV luminosity.}
	\label{tab:limits:jetsplusmet}
\end{table}

The ATLAS and CMS collaborations have both updated their jets + MET searches at $13$~TeV~\cite{Atlas:2016rxq, Khachatryan:2016kdk}, which are not included in the \texttt{CheckMate} catalogue yet. Given the slightly larger luminosity of ATLAS, we recast the ATLAS search only and expect the CMS bounds to be similar.  Using the Monte Carlo pipeline described in the end of section~\ref{sec:pheno:widths}, we cross check our MLM matched sample of signal events with the cut flow efficiencies of the signal regions described in reference~\cite{Atlas:2016rxq}.  We use our derived signal acceptance fractions to compute 95\% C.L.~exclusions on $m_\mathrm{X}$.  Our model is most constrained by the ``2-jet loose'' and ``2-jet medium'' signal regions of reference~\cite{Atlas:2016rxq}, which require at least two hard jets + MET as well as a loose or medium cut on the effective mass.  Our limits on $m_\mathrm{X}$ are shown in the middle row of table~\ref{tab:limits:jetsplusmet}, showing a minor improvement over the 8~TeV results.

In order to compare with the projected sensitivity of our mixed mediator search described in section~\ref{sec:dq3:mixed}, we extrapolate the jets + MET search to $100$~fb$^{-1}$ of integrated luminosity.  Assuming the current systematic uncertainties are unchanged, we rescale the background and signal events by the luminosity ratio for the 2-jet loose signal region to obtain a projected 95\% C.L.~exclusion on the X mass.  Since the 2-jet loose and 2-jet medium signal regions are already dominated by systematic uncertainties, the increase in integrated luminosity does not significantly improve the bounds on the visible cross section and thereby on $m_\mathrm{X}$.  Our results are in the last row of table~\ref{tab:limits:jetsplusmet}, which show that X masses of about 470~GeV are expected to be probed in future multijet + MET searches.

\subsection{Summary of conventional searches}
\label{sec:search:summary}
To summarize, we have studied the existing constraints and future sensitivity from the conventional LHC searches.  We first discussed the dijet resonance searches targeting single and pair production of the mediator, where the bounds are collected in figure~\ref{fig:exclusion:dijet}.  The single production bounds constrain $B \lesssim 35\%$ for $\Gamma_\text{M} / m_\text{M} = 10^{-3}$, although this bound is completely absent once $\Gamma_\text{M} / m_\text{M} = 10^{-4}$.  Current pair production limits constrain mediators below 360~GeV for $B$ varying between about 45\% and 100\%, though this bound is strongest only for mediators near 200~GeV.  The projected reach of the paired dijet search with 100 fb$^{-1}$ of 13~TeV data covers mediator masses between 360~GeV to 720~GeV, but this relies on a simple extrapolation of the low mass and high mass reach from 8~TeV.  We conclude that mediators heavier than 360~GeV with $\Gamma_\text{M} / m_\text{M} \leq 10^{-4}$ are unconstrained by the current single and paired dijet searches, which leaves a large, open parameter space to explore with future dijet searches and our proposed mixed decay search described in section~\ref{sec:dq3:mixed}.

We then discussed the multijet + MET searches, which target X pair production and the X $\to (jj)_\text{soft} \slashed{E}_T$ decay.  Our resulting limits on the X mass, shown in table~\ref{tab:limits:jetsplusmet}, have a mild dependence on $\Delta$, since $\Delta$ sets the hardness of the jets from the X decay.  The jets + MET searches, however, are already dominated by systematic uncertainties, and so they are not expected to improve dramatically with 100 fb$^{-1}$ of 13~TeV data.  Our recasting and projection of the ATLAS 13~TeV search~\cite{Atlas:2016rxq} shows an improvement on the $m_\text{X}$ limit from 400~GeV to 460~GeV.  The results should be juxtaposed with figure~\ref{fig:dq3:relic:density:Gamma}, where the $\Delta = 0.125$ to $\Delta = 0.10$ relic density regions straddle the current and expected exclusion limits from the jets + MET search.  This coincidence in the relic density motivated region indicates that the LHC has excellent prospects for discovering a colored coannihilation mechanism akin to our $s$-channel diquark model, and it also emphasizes the importance of the overlooked dijet resonance + MET channel, which we discuss next.

\section{Mixed decay phenomenology}
\label{sec:dq3:mixed}
As highlighted in section~\ref{sec:pheno:widths}, pair production of our diquark mediator can lead to the unique signature of a dijet resonance accompanied by missing transverse energy.  The resonance + MET final state was also previously emphasized as a general prediction of coannihilation models with an $s$-channel mediator~\cite{Baker:2015qna}.  The resonance + MET signature is also particularly motivated because this channel directly tests the coannihilation diagram from figure~\ref{fig:coannihilationschannel}. Since the rate for this signature is sensitive to the ratio of the dark and visible new physics couplings via the mediator branching fractions, combining results from the mixed and the dijet resonance searches would be central to determining the underlying dark matter Lagrangian.  Moreover, the dijet resonance + MET signature together with the relic density calculations presented in section~\ref{sec:relic:density} and the single dijet, paired dijets, and jets + MET signatures discussed in section~\ref{sec:existing:searches} unify into a compelling picture of a bottom-up discovery of dark matter at the LHC.  This synthesis will be presented in section~\ref{sec:dq3:combination}.

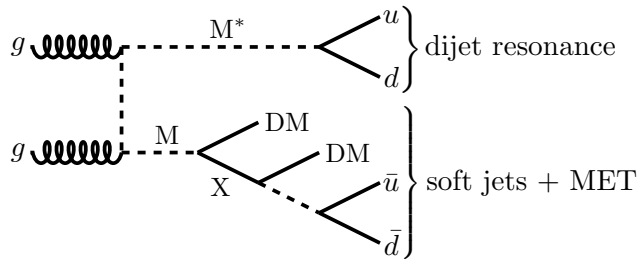
\begin{figure}[tb]
  \centering
  \begin{tikzpicture}[line width=1.4pt, scale=1]
	\draw[gluon] (-1,1.2)--(-2.2,1.2);
	\draw[gluon] (-1,-0.2)--(-2.2,-0.2);
	\draw[scalarna] (-1,1.2)--(-1,-0.2);
	\draw[scalarna] (-1,1.2)--(1.6,1.2);
	\draw[fermionna] (1.6,1.2)--(2.4,1.2+0.4);
	\draw[fermionna] (1.6,1.2)--(2.4,1.2-0.4);
	\draw[scalarna] (-1,-0.2)--(0,-0.2);
	\draw[fermionna] (0,-0.2)--(0.8,0.2);
	\draw[fermionna] (0,-0.2)--(0.8,-0.6);
	\draw[fermionna] (0.8,-0.6)--(1.6,-0.2);
	\draw[scalarna] (0.8,-0.6)--(1.6,-1);
	\draw[fermionna] (1.6,-1)--(2.4,-0.6);
	\draw[fermionna] (1.6,-1)--(2.4,-1.4);
	\node at (-2.36,1.2) {$g$};
	\node at (-2.36,-0.2) {$g$};
	\node at (-0.4,0.03) {\small{M}};
	\node at (0.4,1.45) {\small{$\text{M}^\ast$}};
	\node at (2.55,1.6) {$u$};
	\node at (2.55,0.8) {$d$};
	\node at (1.18,0.2) {\small{DM}};
	\node at (0.3,-0.65) {\small{X}};
	\node at (1.98,-0.2) {\small{DM}};
	\node at (2.55,-0.6) {$\bar{u}$};
	\node at (2.55,-1.4) {$\bar{d}$};
	\node at (2.66,-0.6) {\scriptsize $\left. \begin{array}{c} ~ \\ ~ \\ ~ \\ ~ \\ \end{array} \right\}$};
	\node at (4.40,-0.6) {soft jets + MET};
	\node at (2.8,1.17) {$\Bigg\}$};
	\node at (4.25,1.19) {dijet resonance};
\end{tikzpicture}
  \caption{An example process for the mixed decay signature from pair produced mediators.  One mediator decays visibly to a dijet resonance, and the other decays to DM and X, giving soft jets and missing transverse energy.}
  \label{fig:mixed:signature}
\end{figure}

The mixed decay of pair-produced mediators leads to a final state of $(jj)_{\text{res}} + (jj)_{\text{soft}} + \slashed{E}_T$, as shown in figure~\ref{fig:mixed:signature}.  The typical momenta of the soft jets is determined by the mass splitting between X and DM.  As described in section~\ref{sec:relic:density}, and in most of the parameter space favored by relic density requirements, these momenta are close to the various jet $p_T$ thresholds used by ATLAS and CMS. The primary goal of our analysis is to identify the dijet resonance and large MET signature, while the possible discrimination of additional soft signal jets from X decay will be a minor secondary consideration.

\subsection{Event generation}
\label{sec:dq3:event:generation}

Signal events are generated using the procedure and tools described in section~\ref{sec:model}.  Following the relic density favored regions in figure~\ref{fig:dq3:relic:density:Delta}, we focus on $\Delta = 0.1$, $0.125$, and $0.15$.  Given the jets + MET recasting performed in section~\ref{sec:dq3:jets:met}, with results summarized in table~\ref{tab:limits:jetsplusmet}, we know that X is already excluded to about 400~GeV for these choices of $\Delta$, and hence we focus our study on mediators in the several hundred GeV to TeV mass range.

Our new physics signal is characterized by a strong dijet resonance recoiling against large MET.  While the intrinsic width of our mediator is constrained to be small by dijet resonance searches, as discussed in section~\ref{sec:dq3:dijet}, we expect the resolution on our narrow resonance to be dominated by jet energy scale uncertainty as well as jet clustering contamination from ISR, final state radiation (FSR), and the X decay products.  For our mediator mass range and choices of $\Delta$, we find these effects are not significant enough to warrant a complex jet substructure analysis to sharpen the dijet peak structure, and we instead ameliorate these effects by optimizing the jet clustering algorithm and clustering radius.  Our events are clustered using the anti-$k_T$ algorithm~\cite{Cacciari:2008gp} with $R = 0.5$.

Given the final state of a dijet resonance and MET, we study the following backgrounds:
\begin{itemize}
  \item $(Z\rightarrow \nu \nu) + 1,2~\mathrm{jets}$: this background is similar to the final state that we are considering but will typically have a lower $\slashed{E}_T$ and no resonant structure.
  \item $(W^\pm \rightarrow \ell^\pm \nu) + 1, 2~\mathrm{jets}$: this background can be significantly reduced by a lepton veto.  If the lepton is not reconstructed, though, the associated signature is similar to the $(Z\rightarrow \nu \nu) + \mathrm{jets}$ background.
  \item QCD multijets: while this background has no intrinsic MET, QCD can easily produce fake MET if a jet is badly reconstructed or mismeasured.  This $\slashed{E}_T$ distribution follows a different slope than the vector boson + jets backgrounds above and should lead to a smooth continuum in the dijet invariant mass spectrum.
\end{itemize}
We do not simulate other backgrounds, which are expected to be subdominant to those listed above.  In particular, $t\bar{t} + \text{jets}$ will be readily reduced by vetoing heavy flavor jets, while the remaining semi-leptonic and fully leptonic $t\bar{t}$ events will only contribute if the leptons are lost, which renders it subdominant to the $W^\pm + \text{jets}$ background.  Similarly, diboson production where one boson decays invisibly or leptonically and the other hadronically will be a small correction to the $Z + \text{jets}$ or $W^\pm + \text{jets}$ backgrounds, respectively.  To properly model the associated jet distributions for each background, we use MLM matching~\cite{Mangano:2002ea} to avoid double-counting jets from the hard matrix element and jets from the parton shower.  The $Z$ and $W^\pm$ backgrounds are matched up to two jets and the QCD multijet background is matched up to three jets, each using a matching scale of $20$~GeV.

In order for the Monte Carlo generation to efficiently populate the tails of the background distributions, we sample the phase space following the procedure described in reference~\cite{Avetisyan:2013onh}.  We modify \texttt{MadGraph} to implement an $S_T^*$ cut at generator level, where $S_T^*$ is the scalar $p_T$ sum of all generator level particles. We partition the MC generation in $S_T^*$ bins, which follow
\begin{equation}
  \sigma_i = \sigma(\mathtt{stmin}_i < S_T^* < \mathtt{stmax}_i) \gtrsim 
  0.9 \times \sigma(\mathtt{stmin}_i < S_T^* ) \ ,
\end{equation}
where $\sigma_i$ is the cross section in the $i$-th bin and $\mathtt{stmax}_i$, $\mathtt{stmin}_i$ are the bin edges. The final overflow bin must satisfy $N > 10 \times \sigma_{\mathrm{overflow}} \times L$, where $N$ is the total number of events to be generated in the bin and $L$ is the desired integrated luminosity.

We use \texttt{NLL-fast v3.1}~\cite{Borschensky:2016xy, Beenakker:2009ha, Kulesza:2009kq, Kulesza:2008jb, Beenakker:2015rna} in combination with \texttt{NNPDF v3.0}~\cite{Ball:2014uwa} to calculate $m_\mathrm{M}$ dependent NLO-cross sections for pair-production of the mediator at $13$~TeV.  For our mediator mass range, the $K$-factor is relatively flat, varying between $1.58$ and $1.65$ for $m_\text{M}$ from $500$ to $1500$~GeV.  The background $K$-factors are obtained from \texttt{MCFM v6.8}~\cite{Campbell:2010ff} for processes involving electroweak gauge bosons and from references~\cite{Ellis:1992en, Giele:1994gf} for QCD multijets.  The background $K$-factors are found to be $1.15$ for the vector boson backgrounds and $1.3$ for QCD multijets.

We apply mild preselection cuts on both the signal and backgrounds, requiring events to have at least two jets, $\slashed{E}_T > 100$~GeV and the leading jet $p_T$ greater than $80$~GeV.  These cuts only increase the efficiency of our Monte Carlo event generation and will be superseded by the analysis level cuts.  Signal samples are generated with $m_\text{M}$ starting at $500$~GeV, which is the lowest mass our simulation can confidently probe a dijet resonance, and $m_\text{DM}$ starting at $25$~GeV.\footnote{\texttt{Pythia 8} is not able to correctly process the X decay for the tiny mass splittings associated with lighter DM masses, but in principle, our search strategy will still be sensitive to this region of parameter space.}

\subsection{The mixed decay search strategy}
\label{sec:dq3:search:strategy}
As outlined in~\ref{sec:dq3:event:generation}, our search strategy targets a signal exhibiting a large amount of $\slashed{E}_T$, at least two hard jets with a resonant structure, and possible additional resolved jets from X decays, ISR, and FSR.  Our main discriminants between signal and background will be MET and the resonant dijet peak over a continuum dijet background.  We will also employ angular cuts to improve the overall signal to background ratio in the dijet invariant mass spectrum.

Our cut flow table is shown in table~\ref{tab:mixedcutflow}.  In the first row, we show the 13~TeV cross sections for each process after applying preselection cuts and $K$-factors.  For comparison, we also show a signal benchmark point with $(m_{\mathrm{DM}}, m_{\mathrm{X}}, m_{\mathrm{M}}) = (379, 417, 900)$ GeV.  Sequential cuts and cumulative acceptance efficiencies are shown in subsequent rows.

We first apply a lepton veto, which mainly reduces the leptonic $W^\pm + $ jets background by about $30$\%.  This low suppression rate is due to the tight lepton isolation requirements of the CMS Delphes card (summed $p_T$ of tracks within a cone of $R = 0.5$ must be less than $10$\% of the lepton $p_T$).  We expect a veto on leptons using medium or loose identification criteria would improve the $W^\pm + $ jets rejection and negligibly impact our signal acceptance.  Next, since our diquark mediator is coupled only to first generation quarks, we apply a $b$-jet veto, which is expected to significantly suppress the subdominant $t\bar{t}$ background. We use the $b$-tagging efficiencies quoted in reference~\cite{ATL-PHYS-PUB-2015-022}, with a maximum tagging efficiency of $85$\% and a mistag rate of $0.2$\%.

We require at least two hard jets, where hard jets have $p_T > 100$~GeV and $|\eta| < 2.5$.  This hard jet requirement suppresses the remaining vector boson plus jets backgrounds by about $85$\% and the QCD background by about $68$\% while retaining more than $95$\% of the extant signal.  We also cut on events with $\slashed{E}_T$ arising primarily from jet mismeasurement  by requiring $\Delta \phi(j_i, \slashed{E}_T) > 0.2$ for all hard jets $j_i$.  As expected, this cut significantly reduces QCD, by about a factor of four. We model jet mismeasurements using the default parameters from the CMS card of \texttt{Delphes-3.2.0}. Note that, as mentioned in~\cite{Baker:2015qna}, this modeling leads to an overestimate of the tail of the QCD $\slashed{E}_T$ distribution. Our results can therefore be considered conservative. 

Finally, we require the rapidity difference between the two leading hard jets to be less than $1.3$, which mimics the cut used in the dijet resonance searches and reflects the kinematic configuration expected from figure~\ref{fig:mixed:signature}.  After these general cuts, the cumulative acceptances for the different backgrounds are of the order of $5\%$.  We obtain our final sensitivity estimates after simultaneously optimizing the $\slashed{E}_T$ and $m_{j_1 j_2}$ mass window cuts for each signal benchmark.

\begin{table}[tb]
  \centering
  \scriptsize
  \begin{tabular}{!{\vrule width 1pt}c !{\vrule width 1pt} ccccc!{\vrule width 1pt}}
	\noalign{\hrule height 1pt}
	& signal &$Z_{\nu\nu}$ + jets   & $W^+_{\ell\nu}$ + jets  &  $W^- _{\ell\nu}$ +  jets & QCD \\
	$\sigma_{\text{NLO}}$ (pb) & $7.60\! \times\! 10^{-3}$ & $93.21$ & $157.33$ & $75.48$ & $6948.76$\\
	\noalign{\hrule height 1pt}
	$\ell$ veto & $100$ &  $100$ & $71.4$ & $69.0$ & $100$\\
	$b$-jet veto &  $86.9$ & $89.3$ & $66.8$ & $64.5$ & $87.3$ \\
	$p_{T} (j_1),\ \! p_{T}(j_2)\! >\! 100$~GeV& $83.0$   & $11.8$ & $9.76$ & $9.68$ & $27.6$\\
	$\Delta \phi (j_i, \slashed{E}_T)\! >\! 0.2$ & $75.0$ & $10.9$ & $8.50$ & $7.99$ & $6.20$\\
	$\left|\Delta\eta(j_1, j_2)\right|\! <\! 1.3$ & $51.5$ & $6.36$ & $5.01$ & $4.82$ & $3.61$\\
	\noalign{\hrule height 0.5pt}
	$\slashed{E}_T\! >\! 900~\mathrm{GeV}$ & $6.29~\!(48)$ &$3.51\!\times\! 10^{-3}~\!(327)$ & $7.74\!\times\! 10^{-4}~\!(122)$ & $2.83\!\times\! 10^{-4}~\!(21)$ & $1.71\!\times\! 10^{-5}~\!(119)$\\
	Mass window & $3.56~\!(27)$ &  $3.35\! \times\! 10^{-4}~\!(31)$ & $8.70\!\times\! 10^{-5}~\!(14)$ & $1.93\!\times\! 10^{-5}~\!(2)$ & $1.28\!\times\! 10^{-6}~\!(9)$\\
	\noalign{\hrule height 1pt}
\end{tabular}

  \caption{Detailed cut flow for a signal benchmark with $(m_{\mathrm{DM}},m_{\mathrm{X}},m_{\mathrm{M}})$ = (379, 417, 900)~GeV and the $Z + \text{jets}$, $W^\pm + \text{jets}$, and QCD multijet backgrounds.  The cross sections quoted include the preliminary cuts described in section~\ref{sec:dq3:event:generation} and $K$-factors for 13~TeV LHC.  The $\slashed{E}_T$ and $m_{j_1j_2}$ cuts are optimized for each signal point individually, and we quote both the efficiency (in \%) and the number of events at $100$~fb$^{-1}$ luminosity for each process.  The dijet mass window considered for this benchmark point is $\left[ 812, 987 \right]$~GeV.}
  \label{tab:mixedcutflow}
\end{table}

\begin{figure}[tb]
  \centering
  \includegraphics[width=\textwidth]{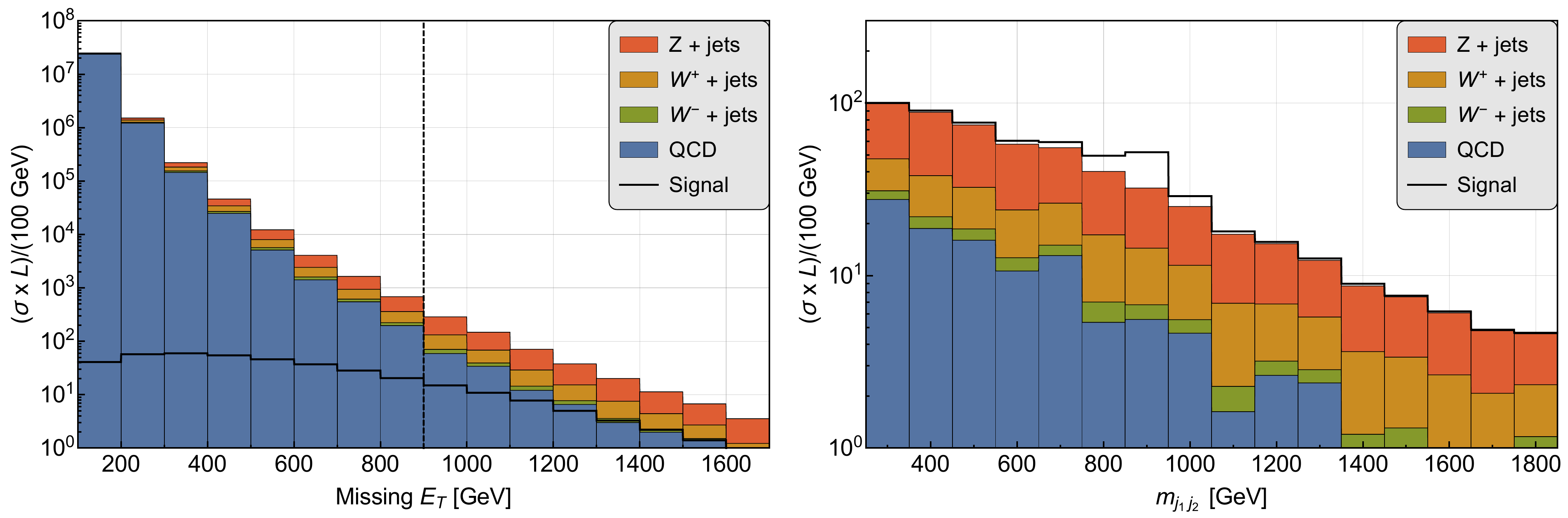}
  \caption{Signal and background $\slashed{E}_T$ (left) and $m_{j_1 j_2}$ (right) distributions for a signal benchmark model with $(m_{\mathrm{DM}},m_{\mathrm{X}},m_{\mathrm{M}})$ = (379, 417, 900)~GeV.  The signal distribution is shown in black outline.  The stacked background distributions are from top to bottom: $Z + $ jets (red), $W^+ + $ jets (orange), $W^- + $ jets (green) and QCD (blue).  The vertical dashed line at $900$~GeV on the left panel represents the optimal $\slashed{E}_T$ cut for this benchmark point.  The $m_{j_1j_2}$ distribution is shown for events that passed this $\slashed{E}_T$ cut.}
  \label{fig:mixed:distributions}
\end{figure}

In the left panel of figure~\ref{fig:mixed:distributions}, we show the stacked $\slashed{E}_T$ distribution after cuts for the different backgrounds, and overlay our signal benchmark as a black outline.  Cutting on $\slashed{E}_T$ significantly removes the backgrounds, where the rejection factors range from $10^3$ for $Z + $ jets to $10^5$ for QCD, while the benchmark signal is reduced by less than an order of magnitude, giving a signal to background ratio of $\mathcal{O}(10\%)$.  The optimal $\slashed{E}_T$ cut is correlated with the mediator mass.  Furthermore, for low DM masses, jets from the X decay might be boosted enough to pass the detection threshold, thus the amount of $\slashed{E}_T$ in signal events is also expected to mildly decrease with the DM mass.  Note that, at this point of the search, our analysis strategy is similar to the one used in the existing multijet + MET searches, but the resonant structure of the mediator provides a significant handle to dramatically improve the mediator mass reach.

In the right panel of figure~\ref{fig:mixed:distributions}, we show the $m_{j_1j_2}$ distributions for the signal and the backgrounds after a $\slashed{E}_T > 900$~GeV cut.  Since the background jets do not come from resonances, the background dijet mass spectrum falls smoothly.  The signal, however, exhibits a wide peak around the mass of the mediator particle and can be easily identified on top of the backgrounds.  To impose an appropriate mass window for the dijet mass spectrum, we fit the signal $m_{j_1 j_2}$ distribution after cuts by a Crystal Ball function~\cite{Gaiser:1982yw},
\begin{equation}
  s(x) = N\times 
  \begin{cases}
    \exp \left[ -\frac{(x - \bar x)^2}{2\sigma^2} \right] 
\text{ if } \frac{x - \bar x}{\sigma} > -\alpha \ , \\
    \left( \frac{n}{|\alpha|} \right)^n 
    \exp \left( -\frac{|\alpha|^2}{2} \right) \left( \frac{n}{|\alpha|} 
- |\alpha| - \frac{x-\bar x}{\sigma} \right)^{-n} 
\text{ if } \frac{x - \bar x}{\sigma} \leq -\alpha \ .
  \end{cases}
\end{equation}
Here $N$ is the normalization, $\bar x$ and $\sigma$ are Gaussian parameters, and $\alpha$ and $n$ are power law parameters, all determined by the fit.  We integrate the number of events within $2\sigma$ of the Gaussian mean, $\bar{x}$, for the signal and the background samples to calculate the significance,
\begin{equation}
  \mathcal{S} = 
  \frac{S}{\sqrt{\sum B_i + (\sum \mathrm{sys}_i \times B_i)^2}} \ ,
\end{equation}
where $\mathrm{sys}_i$ is the systematic uncertainty for a given background $B_i$, and $S$ and $B_i$ are the number of signal and background events in the mass window, respectively.  Following reference~\cite{Atlas:2016rxq}, we take $\mathrm{sys}_i = 5\%$ for the $W$ and $Z$ backgrounds, while we use $20\%$ for the QCD multijet background, because of the lack of simulated events in the tail of the $\slashed{E}_T$ distribution.

We optimize the $\slashed{E}_T$ cut by maximizing the dijet resonance significance $\mathcal{S}$.  The last two rows of table~\ref{tab:mixedcutflow} show the efficiencies as well as expected number of events for 100 fb$^{-1}$ of 13~TeV LHC data given the benchmark-specific $\slashed{E}_T > 900$~GeV cut and the dijet mass window cut, $812 \text{ GeV } \leq m_{j_1 j_2} \leq 987 \text{ GeV}$.

In order to estimate robustly the number of background events in the mass window after the $\slashed{E}_T$ cut without being limited by Monte Carlo statistics, we fit each background $m_{j_1 j_2}$ distribution by a falling function of the form~\cite{Aad:2014aqa}
\begin{equation}
  b(x) = p_0 (1 - x)^{p_1} x^{p_2 + p_3 \ln x} \ ,
\label{eq:bfit}
\end{equation}
where $x = m_{j_1 j_2} / \sqrt{s}$. For each background, we then integrate $b(x)$ with the optimal $p_i$ parameters over the chosen mass window to obtain the number of background events after the $m_{j_1j_2}$ cuts.

\subsection{Sensitivity of the mixed decay search}
\label{sec:dq3:hunting}

\begin{figure}[tb]
  \centering
  \includegraphics[width=0.8\textwidth]{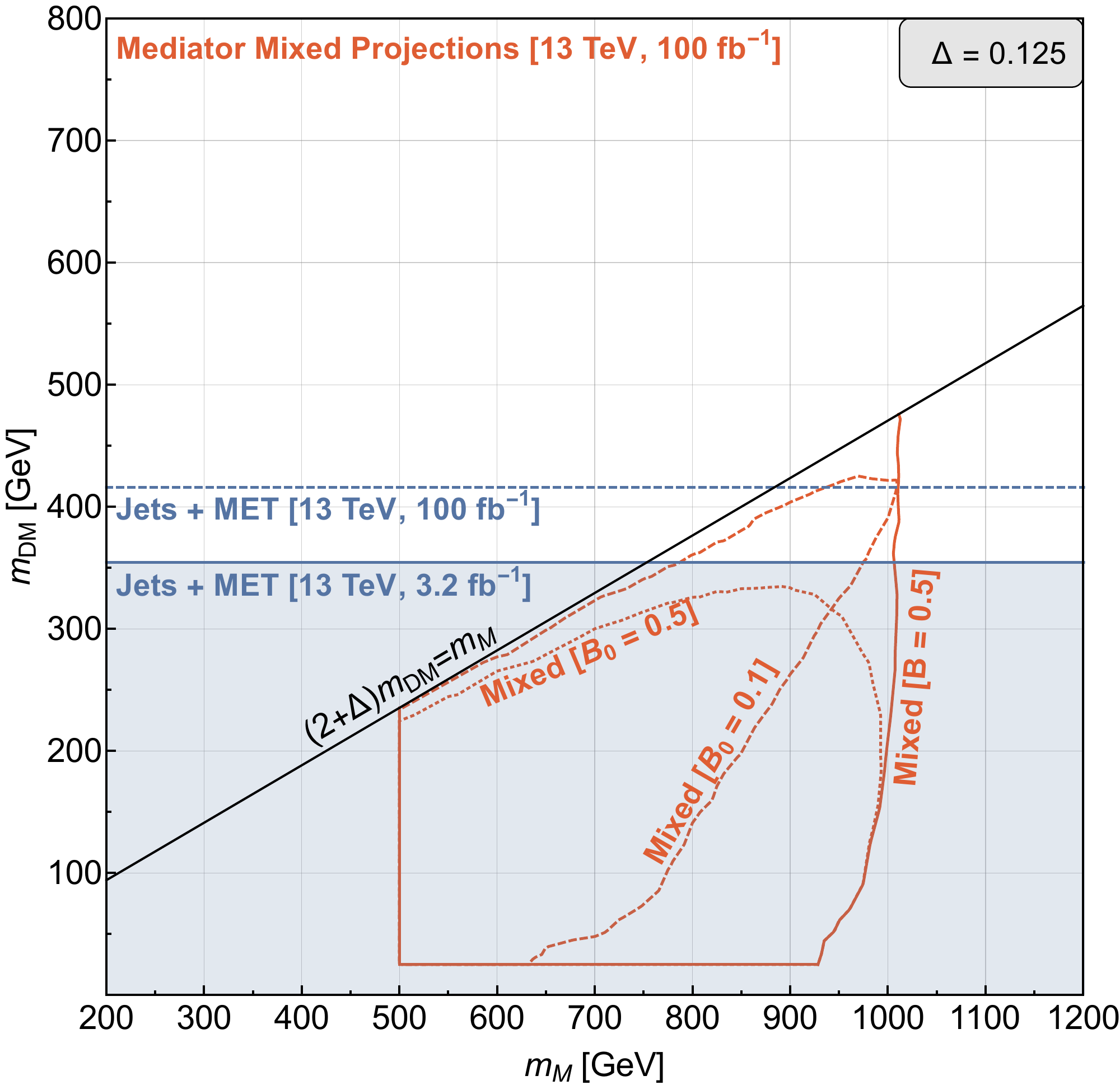}
  \caption{Comparison in LHC sensitivity between the existing jets + MET searches and the newly proposed mixed topology in the DM mass vs.~mediator mass plane with $\Delta = 0.125$.  The mixed search sensitivity is shown for $B_0 = 0.1$, $B_0 = 0.5$, and $B = \text{Br} \left( \text{M} \to \bar{u} \bar{d} \right) = 0.5$ as an ideal scenario.  For given values of $m_\mathrm{DM}$ and $m_\mathrm{M}$, this ideal rate can always be ensured by choosing $B_0$ accordingly.}
  \label{fig:exclusion:projection:maximum}
\end{figure}

The results of our collider study for $\Delta = 0.125$ are shown in figure~\ref{fig:exclusion:projection:maximum} for $100$~fb$^{-1}$ of 13~TeV luminosity.  We illustrate the mixed sensitivity with three contours corresponding to $B_0 = 0.1$, $B_0 = 0.5$, and $B = 0.5$.  Recall that fixing $B_0$ is equivalent to fixing a particular coupling combination of $y_{ud}$ and $y_D$ via equation~\ref{eq:B0:definition}, while the physical branching fraction of the mediator to dijets depends on $B_0$ and the phase space of the M $\to$ X DM decay.  Alternately, fixing $B$ requires the couplings to change as a function of the M and DM masses.  In particular, as we move along the $B = 0.5$ contour in figure~\ref{fig:exclusion:projection:maximum}, $B_0$ is monotonically decreasing to zero up to the phase space boundary of $(2 + \Delta) m_\text{DM} = m_\text{M}$.

As expected, by changing from $B_0 = 0.5$ to $B_0 = 0.1$, the expected reach of the dijet resonance + MET search moves closer to the phase space boundary of $(2 + \Delta) m_\text{DM} = m_\text{M}$.  The $B_0 = 0.5$ reach is coincident with the $B = 0.5$ reach for low DM masses since the phase space suppression of $K(\Delta, \tau_\text{DM})$ is negligible.  By construction, the intersection point between the $B_0 = 0.1$ and $B = 0.5$ contours indicates when the phase space suppression factor maximizes the overall rate for the mixed decay signature from M pair production, $\sigma (p p \to \text{M}^* \text{M}) \times 2B (1-B) \xrightarrow[B = 0.5]{} 0.5 \times \sigma (p p \to \text{M}^* \text{M})$.  The $B = 0.5$ choice also fixes the paired dijet resonance rate to $0.25 \times \sigma (p p \to \text{M}^* \text{M})$ when the M $\to$ X DM decay is on-shell.  The $B_0 = 0.1$ and $B = 0.5$ contours indicate that a significant portion of parameter space in the $m_\text{DM}, m_\text{M}$ plane will be tested by the mixed decay search from section~\ref{sec:dq3:search:strategy}.  This search is also complementary to the current and future reach from the jets + MET searches for X pair production, indicated by the shaded and horizontal lines at $m_\text{DM} \sim 360$~GeV and $m_\text{DM} \sim 420$~GeV in figure~\ref{fig:exclusion:projection:maximum}, respectively.  These two searches test integral processes predicted by the coannihilation mechanism, and the simultaneous observation of both channels would be a striking sign of $s$-channel coannihilation at the LHC.

\subsection{Connecting the relic density to collider searches}
\label{sec:dq3:combination}

As emphasized in reference~\cite{Baker:2015qna} and throughout this paper, our simplified models based on modeling the DM annihilation mechanism allow us to develop collider searches that target the parameter space favored by the measured $\Omega h^2$.  Moreover, for our model, direct detection and indrect detection signals are suppressed compared to the collider probes, and thus the relic density parameter space will mainly be confronted by the existing LHC searches discussed in section~\ref{sec:existing:searches} and our proposed mixed decay analysis from section~\ref{sec:dq3:search:strategy}.  This bottom-up approach directly connects the relic density requirement to the LHC signature space, enforcing the role of LHC as a DM producing machine.

As discussed in section~\ref{sec:relic:density}, DM (co)annihilation to SM particles occurs through numerous subprocesses with different dependencies in the new physics couplings $y_{ud}$ and $y_D$.  These couplings, though, are significantly constrained by the dijet resonance searches described in section~\ref{sec:dq3:dijet}.  As a result, in the parameter space region where DM has the correct relic density and satisfies the dijet constraints, the relic density will be driven by $\overline{\text{X}}$~X annihilation via strong interactions or DM X resonant coannihilation.  This was illustrated in figure~\ref{fig:dq3:relic:density:Delta}, where the chosen $B_0 = 0.1$ parameter and $\Gamma_\text{M} / m_\text{M} \leq 10^{-4}$ scan led to $\Omega h^2$ favored regions for each value of $\Delta$.  The relic density analysis points to $\Delta \sim 10\%$ for DM and mediator masses within reach of the LHC, although the exact preferred region depends sensitively on $\Delta$, as shown in figure~\ref{fig:dq3:relic:density:Delta}.

\begin{figure}[tb]
  \centering
  \includegraphics[width=0.8\textwidth]{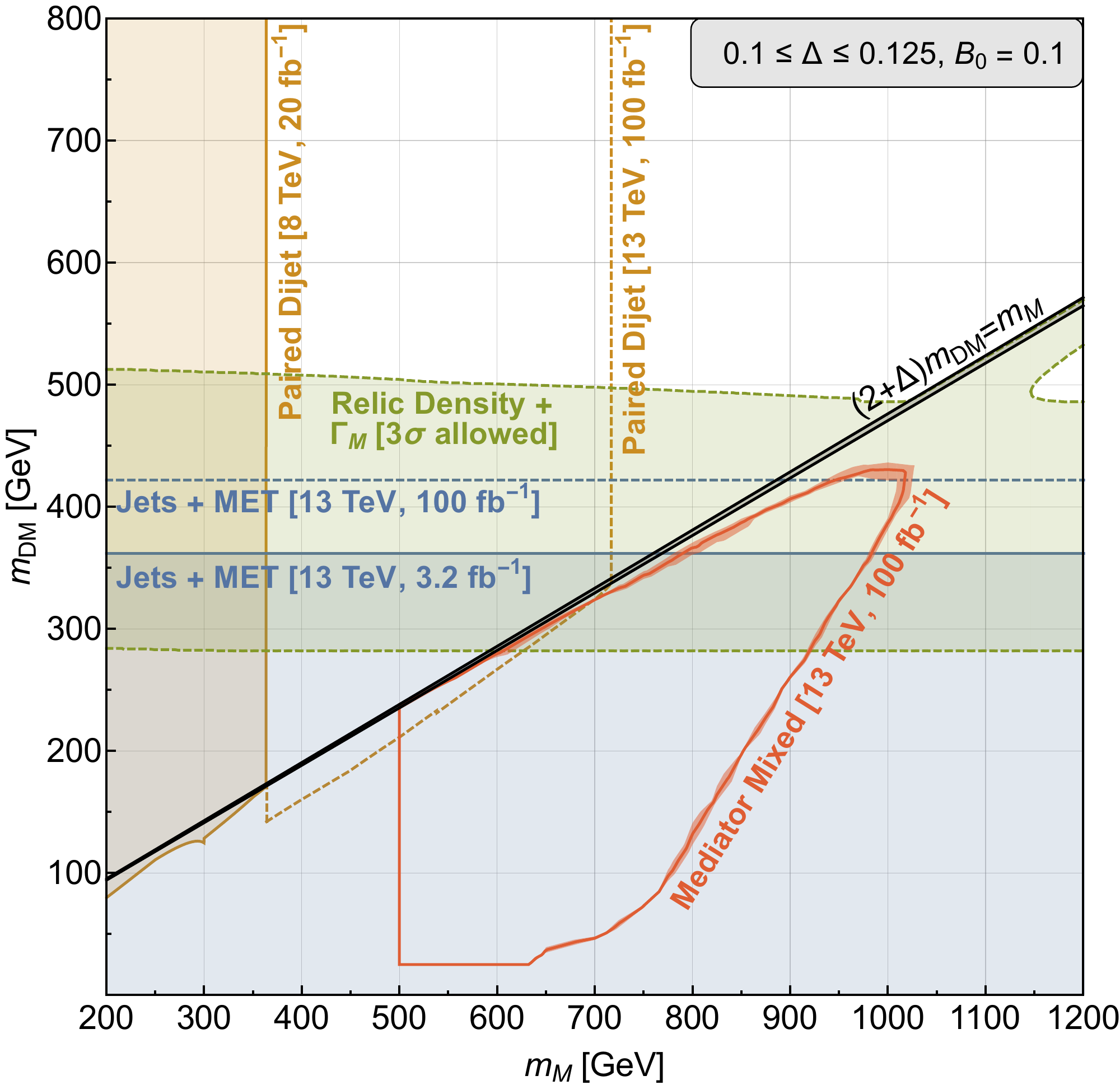}
  \caption{Summary plot for $0.1 \leq \Delta \leq 0.125$ and $B_0 = 0.1$ in the dark matter mass versus mediator mass plane.  The relic density region is determined by scanning over mediator couplings requiring $\Gamma_\text{M} / m_\text{M} \leq 10^{-4}$ and matching the Planck measurement of $\Omega h^2$, allowing a $3\sigma$ deviation from the central value.  The current 13~TeV, 3.2 fb$^{-1}$ limit from recasting the ATLAS jets + MET search~\cite{Atlas:2016rxq} and the projection with 100 fb$^{-1}$ luminosity are shown in solid and dashed blue lines, adopting the strongest limit according to the variation in $\Delta$.  The current 8~TeV, 20 fb$^{-1}$ and future 13~TeV, 100 fb$^{-1}$ paired dijet limits are shown as mostly vertical orange contours.  Our proposed search for the novel signature of a dijet resonance + MET is shown for 100 fb$^{-1}$ of 13~TeV data as a red contour with shading to indicate the dependence on $\Delta$.  The black diagonal lines span the kinematic threshold $(2 + \Delta) m_\text{DM} = m_\text{M}$ for on-shell decays of the mediator to X DM.}
  \label{fig:exclusion:projection:10:125}
\end{figure}

The prospects of different LHC searches as well as the region allowed by relic density constraints are shown in figure~\ref{fig:exclusion:projection:10:125} for $0.10 \leq \Delta \leq 0.125$.  For this $\Delta$ range, the correct relic density is obtained for dark matter masses between $300$ to $500$~GeV or in the funnel region.  Following the discussion in section~\ref{sec:relic:density:regions}, the relic density region constricts to a horizontal band driven by $\overline{\text{X}}$ X $\to$ $\overline{\text{SM}}$ SM and a coannihilation funnel parallel to the $(2 + \Delta) m_{\text{DM}} = m_\text{M}$ line.  Furthermore, $\Delta = 0.125$ corresponds to the lower edge of the horizontal band, while $\Delta = 0.1$ corresponds to the upper edge.

While $m_\text{DM} \lesssim 360$~GeV is excluded by the recasted 13~TeV jets + MET search by ATLAS discussed in section~\ref{sec:dq3:jets:met}, the projected limit of this search with 100 fb$^{-1}$ luminosity is only expected to improve to $m_\text{DM} \sim 425$~GeV and will only cover part of the favored relic density region.  Similarly, the relic density region is excluded for $m_\text{M} \lesssim 360$~GeV by the 8~TeV search for paired dijet resonances, but this search weakens dramatically below the kinematic threshold $(2 + \Delta) m_\text{DM} = m_\text{M}$ because of the rapid turn-on of the M $\to$ X DM decay for our choice of $B_0 = 0.1$.  Nevertheless, the 100 fb$^{-1}$ projection for paired dijet searches should reach mediator masses of about 720~GeV, giving an independent test of the relic density region above the $(2 + \Delta) m_\text{DM} = m_\text{M}$ line compared to the 100 fb$^{-1}$ jets + MET projection.

Moreover, for on-shell decays of the mediator to X DM, the new mediator mixed decay search introduced in sections~\ref{sec:dq3:search:strategy} and~\ref{sec:dq3:hunting} probes an orthogonal decay mode to the paired dijet search and also complements and enhances the jets + MET reach.  We remark that fixing $B = 0.5$ gave a uniform reach for the mediator mixed search at about $m_\text{M} = 1$~TeV, as shown in figure~\ref{fig:exclusion:projection:maximum}, and so the mixed decay search clearly has parameter space sensitivity beyond that expected from the jets + MET projection.  In the fortunate case of excesses in both the mixed decay search and the jets + MET search, an immediate goal would be to test whether the mediator is consistent with an $s$-channel coannihilation mechanism.  Multiple, well-motivated searches covering complementary signatures of the same relic density region is a hallmark feature of simplified models in our coannihilation  framework, since these signatures result from recycling vertices and gauge number assignments inherent in the DM coannihilation diagram.

\section{Conclusions}
\label{sec:conclusion}

The enigmatic dark matter question remains at the forefront of
particle physics research today.  Discovering the particle properties
of dark matter would represent a pinnacle achievement in the fields of
particle physics and astrophysics.  Hence it is crucially important to
identify all viable dark matter search strategies for present and
future experiments.

Building on the general classification of simplified coannihilating
dark matter models~\cite{Baker:2015qna}, we studied the relic density
predictions, indirect and direct detection constraints, and collider
phenomenology of a dark matter particle coannihilating with a strongly
interacting partner.  We adopted model ST6 from
reference~\cite{Baker:2015qna} as a case study, where the DM is an
electroweak singlet Majorana fermion, the coannihilation partner X is
a color triplet fermion, and the $s$-channel mediator M is a scalar
color triplet.  Besides the three masses, the only other relevant
parameters in this simplified model are the $y_D$ coupling between the
mediator, X, and DM, and the $y_{ud}$ coupling between the mediator
and the SM up and down quarks, ensuring consistency with flavor
constraints and precision Higgs physics.

Dark matter models with a strongly interacting coannihilation partner
have two striking features regarding relic density and collider
signatures.  First, the coannihilation paradigm provides an explicit
counterpart to the WIMP miracle, giving weak scale masses and the
correct relic density unencumbered by weak couplings or weak scale
cross sections.  In particular, the relic density is determined by
four types of (co)annihilation channels.  The DM X $\to$ SM$_1$ SM$_2$
channel dominates in the funnel region, where $m_\mathrm{DM} +
m_\mathrm{X} \approx m_\mathrm{M}$.  The DM X $\to$ M $V_\mathrm{SM}$
and $\overline{\text{DM}}$ DM $\to$ M$^*$ M processes can become
relevant for large DM masses, above the resonant coannihilation
funnel.  The $\overline{\text{X}}$ X $\to$ $\overline{\text{SM}}$ SM
process, since X has a color charge, efficiently annihilates X
particles into quark and gluon pairs, and is most important in the low
$m_\mathrm{DM}$ region.  This process determines the underclosure bound on
$m_\mathrm{DM}$ for a given fractional X--DM mass splitting $\Delta$
and can push the preferred dark matter mass into the multi-TeV range for
$\Delta\lesssim 0.01$.  For $0.1 \leq \Delta \leq 0.125$, though,
the relic density motivates dark matter masses $m_\mathrm{DM} \sim
300-500$ GeV, consistent with direct and indirect detection
constraints and within the current and near future reach of LHC
searches.

The existence of a colored coannihilation partner and a colored
mediator M ensures strong pair production of these particles at hadron
colliders, and since M couples to first generation quarks, single
mediator production is also possible.  These production modes, when
considered in tandem with the decay processes dictated by the
$s$-channel coannihilation diagram, present very promising and
interesting targets for LHC searches.

The conventional searches at ATLAS and CMS already provide interesting
constraints on the parameter space of the model.  We have seen that
single dijet searches constrain a combination of the mediator width
and its dark and visible couplings, which in turn reduces the
importance of the DM X $\to$ M $V_\text{SM}$ and
$\overline{\text{DM}}$ DM $\to$ M$^*$ M subprocesses in the relic
density calculation and narrows the width of the coannihilation
funnel.  We have also seen that X pair production is probed in jets +
MET searches, where, for $0.1 \leq \Delta \leq 0.125$, X masses below
about 400~GeV are currently ruled out, and X masses around 470~GeV
should be probed with 100 fb$^{-1}$ of 13~TeV data.  In addition, M
pair production gives rise to a paired dijet resonance signature,
where the current searches constrain M masses below 360~GeV and can
potentially reach 720~GeV, as long as the visible branching fraction
is 100\%.

Once the M $\to$ X DM decay is kinematically open, however, our
proposed mediator mixed decay search pushes the mediator mass
sensitivity to roughly 1~TeV.  This can surpass the projected reach of
the jets + MET search in covering the parameter space favored by the
relic density calculation.  This dijet resonance plus MET signal is
characteristic of many $s$-channel coannihilation models featuring a
colored mediator.  Moreover, this signature provides the only direct
link between visible matter and the dark matter, and measurements of
the rates and kinematics in this final state would be crucial for a
determination of the dark matter Lagrangian.

Finally, we stress the intricate interplay between the relic density
analysis, the fractional mass splitting parameter $\Delta$, and the
resulting reach by LHC searches, which are entirely driven the
coannihilation simplified model.  The minimal assumptions of the
coannihilation framework~\cite{Baker:2015qna} suffice to characterize
completely the dark matter annihilation process, ensuring that the
resulting collider signatures are directly motivated and driven by the
potential discovery of thermal dark matter.

\acknowledgments
We would like to thank Mihailo Backovic, Alexander Pukhov, Peter Skands and Stefan Vogl for useful correspondence.  This research is supported by the Cluster of Excellence Precision Physics, Fundamental Interactions and Structure of Matter (PRISMA-EXC 1098), by the ERC Advanced Grant EFT4LHC of the European Research Council, and by the Mainz Institute for Theoretical Physics.  The work of MB and JL is supported by the German Research Foundation (DFG) in the framework of the Research Unit ``New Physics at the Large Hadron Collider'' (FOR 2239) and of Grant No.~KO 4820/1–1.  MB would furthermore like to thank the Fermilab Theory Group for their hospitality.  SEH would also like to acknowledge the hospitality of the Berkeley Center for Theoretical Physics and the Theoretical Physics Department at LBNL.

\bibliographystyle{JHEP}
\bibliography{dijet_met}

\end{document}